\documentclass[12pt,letterpaper,oneside,journal=jacsat]{article}
\usepackage[utf8]{inputenc}
\usepackage{amsmath}
\usepackage{subcaption}
\usepackage{amsfonts}
\usepackage{times}
\usepackage{graphicx}
\usepackage{bm}
\usepackage{achemso}
\usepackage{color}
\usepackage{soul}
\usepackage[table]{xcolor}
\usepackage{authblk}
\usepackage[hidelinks]{hyperref}
\usepackage{tocloft}
\usepackage{tabularx,booktabs}
\usepackage{multirow}
\usepackage[capitalize]{cleveref}
\usepackage{setspace}
\usepackage[margin=1.0 in]{geometry}
\usepackage{indentfirst}
\usepackage{float}
\usepackage{xr}
\setkeys{acs}{articletitle=true}
\setkeys{acs}{maxauthors = 0}

\newcommand{\tc}{$T_\text{c}$}

\DeclareUnicodeCharacter{202F}{\,}
\DeclareUnicodeCharacter{2009}{\,}

\newcolumntype{M}[1]{>{\centering\arraybackslash}m{#1}}

\title{Metal Atom (Dis)Order and Superconductivity in YCaH$_{n}$ ($n=8-20$) High-Pressure Superhydrides}
\author[1]{Masashi W. Kimura}
\author[]{Seong Won Jang}
\author[]{Nisha Geng}
\author[1]{Eva Zurek
 \thanks{Corresponding Author: ezurek@buffalo.edu}}
\affil[1]{Department of Chemistry, State University of New York at Buffalo, Buffalo, NY 14260-3000, USA}

\date{\today}

\begin{document}
\baselineskip24pt

\maketitle
\normalsize

\clearpage
\newpage

\begin{abstract}
High-pressure superhydrides have attracted much attention due to their high superconducting critical temperatures (\tc s). Herein, density functional theory (DFT) calculations are used to study the structures and properties, including potential for metal atom disorder and doping-enhanced \tc , within Y-Ca superhydrides with YCaH$_{n}$ ($n=8-20$) compositions. For YCaH$_8$ numerous phases that differed in the arrangement of the metal atoms were found to be nearly isoenthalpic, suggesting the importance of configurational entropy on stability. The equimolar ratio of the two metal atoms brought the Fermi level to a peak in the density of states, enhancing \tc\ to 149~K and 170~K for  $P4/mmm$ and $Cmmm$ YCaH$_{8}$, respectively, at 180~GPa within the isotropic Eliashberg formalism. YCaH$_{12}$ was also predicted to be disordered, however the \tc s of the ordered variants spanned a wide range from 105-253~K at 200~GPa, showing that doping could either mildly enhance or drastically reduce \tc\ from that of the parent compounds. For YCaH$_{18}$ and YCaH$_{20}$, only a single dynamically stable superhydride was predicted, which we attribute to the differences in the structures of the stable binary parents.
\end{abstract}

\clearpage
\newpage

\section{Introduction}
A decade has passed since the first experiments on a superconducting high-pressure hydride were reported. The record breaking superconducting critical temperature, \tc\ , measured by Eremets' team, 203~K at 150~GPa -- a pressure that is roughly half that found in the Earth's core -- was attributed to a cubic H$_3$S phase where each hydrogen atom is coordinated to two sulfur atoms, and each sulfur atom to six hydrogen atoms~\cite{Drozdov:2015a}. This record was soon surpassed when a \tc\ of 260~K at 180-200~GPa was measured in a compound characterized as LaH$_{10}$~\cite{Drozdov:2019a,Zulu:2019a}. In this phase, each hydrogen atom is weakly bonded to four surrounding hydrogens forming a cage within which the electropositive metal resides. Since then numerous high pressure hydride superconductors have been synthesized, with those containing rare earth or alkaline earth metals possessing the highest measured \tc s, $e.g.$, YH$_{4}$ (\tc\ = 88~K~@155~GPa or 82~K at 167~GPa)~\cite{Shao:2021a,Wang:2022a}, YH$_{6}$ (\tc\ = 218~K @ 165~GPa, 224~K~@ 166~GPa, 220~K~@ 183~GPa)~\cite{Wang:2022a,Troyan:2021a,Kong:2021a}, YH$_{9}$ (\tc\ = 243~K~@ 201~GPa)\cite{Kong:2021a}, CaH$_{6}$ (\tc\ = 215~K~@ 172~GPa or 210~K~@ 160~GPa)~\cite{Ma:2022b,Li:CaH6},  ThH$_{9}$ (\tc\ = 146~K~@ 170-175~GPa)\cite{Semenok:2020a}, ThH$_{10}$ (\tc\ = 161~K~@170-175~GPa)~\cite{Semenok:2020a}, CeH$_{9}$ (\tc\ = $\sim$100~K~@ 130~GPa)~\cite{Chen:2021a} and CeH$_{10}$ (\tc\ = $\sim$115~K~@ 95~GPa)~\cite{Chen:2021a}. The simultaneous measurement of two experimental signatures of superconductivity -- the Meissner effect and a drop of the resistance -- has been reported in CeH$_9$~\cite{Bhattacharyya:2024a}.  

\begin{figure}[!htb]
    \centering
    \includegraphics[width=16 cm]{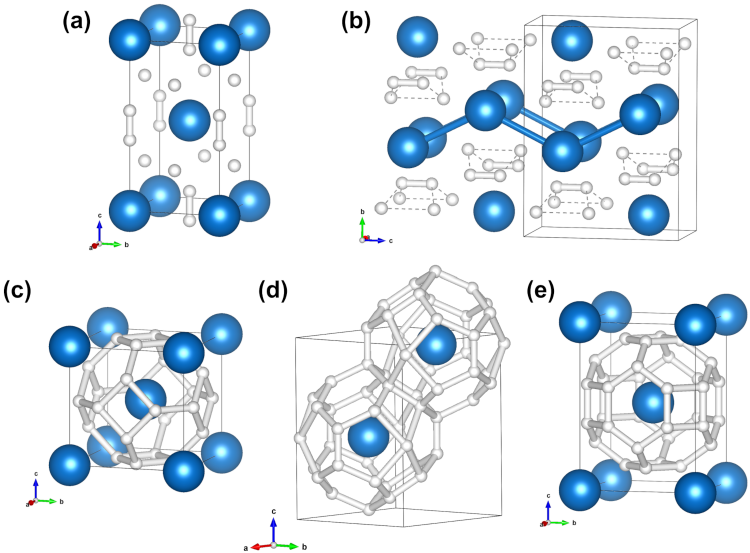}
\caption{The crystal structures of the (a) $I4/mmm$ MH$_4$, (b) $Cmcm$ MH$_{4}$, (c) $Im\bar{3}m$ MH$_6$, (d) $P6_3/mmc$ MH$_9$ and (e) $Fm\bar{3}m$ MH$_{10}$ phases, which are characteristic of many of the predicted and synthesized high-$T_\text{c}$ high pressure binary hydrides. Metal and hydrogen atoms are blue and white, respectively.}\label{fig:strucs}
\end{figure}

Superconductivity in binary high pressure hydrides of an electropositive metal was first predicted using density functional theory (DFT) aided crystal structure prediction (CSP) searches~\cite{Zurek:2021k,Zurek:2022k,Livas:2020-Review,Sun:2024rev}.  These studies have shown that a handful of structure-types are consistently found to be stable, independent of the identity of the metal atom. This includes $I4/mmm$ symmetry MH$_4$ compounds (Figure~\ref{fig:strucs}(a)), which resemble the iconic BaAl$_4$ structure type. These metal tetrahydrides possess two basal hydridic (H$^-$) atoms, and two apical hydrogen atoms that can bond with each other forming H$_2$ molecular units (depending upon the oxidation state, pressure, and radius of the metal atom)~\cite{Bi:2021a}. $Cmcm$ symmetry MH$_{4}$-based compounds have also been predicted~\cite{Wang:2012a,Hooper:2014} (Figure \ref{fig:strucs}(b)). Buckled sheets formed between the nearest-neighbor metal atoms, which interweave perpendicularly through zigzag chains composed of either square arrangements of H$^-$ atoms or distorted polygons of H$^-$ and H$_{2}$ units reminiscent of isosceles trapezoids, comprise these tetrahydrides~\cite{Hooper:2014}. Numerous MH$_6$ compounds have been predicted to possess a sodalite-like H$_{24}$ clathrate cage composed of six square and eight hexagonal faces (Figure~\ref{fig:strucs}(c)), first reported in $Im\bar{3}m$ CaH$_{6}$~\cite{Wang:2012a}. MH$_9$ compositions, on the other hand, often contain an H$_{29}$ cage with six pentagonal, six distorted square, and six hexagonal faces (Figure~\ref{fig:strucs}(d)); while H$_{32}$ cages with six square and twelve hexagonal faces comprise the MH$_{10}$ superhydrides (Figure~\ref{fig:strucs}(e))~\cite{Peng:2017a}. In all of these structure types, the electropositive elements donate electrons to the hydrogenic lattice, resulting in weakened H-H bonds, the dissociation of molecular hydrogen into monoatomic hydrides, or the formation of higher-dimensional hydrogenic lattices featuring multi-centered H-H bonds~\cite{Wang:2012a,Zurek:2019a}.

Moving to ternary and quaternary hydrides has been motivated by the allure of higher \tc s, such as in predicted `hot' superconductors, many of which are based upon more complex clathrate hydrogenic cages. This includes an $Fd\bar{3}m$ Li$_{2}$MgH$_{16}$ phase, which was predicted to have an impressive \tc\ of 473~K at 250~GPa~\cite{Sun:2019}, and an isotypic Li$_{2}$CaH$_{16}$ structure with a \tc\ as high as 330~K at 350~GPa~\cite{Zurek:2024f}. Other examples include LaSc$_{2}$H$_{24}$ (\tc\ = 316~K~@ 167~GPa)~\cite{He:2024a} and YSc$_{2}$H$_{24}$ (\tc\ = 330~K~@ 310~GPa)~\cite{Pham:2024a}, both possessing H$_{24}$ and H$_{30}$ clathrate cages that differ from the MH$_{6}$ and MH$_{10}$ structure types shown in Figures \ref{fig:strucs}c and \ref{fig:strucs}e, respectively. Another goal has been to study hydrides that remain stable and superconducting at lower pressures, such as predicted LaBH$_{8}$~\cite{DiCataldo:2021a} and synthesized LaBeH$_{8}$~\cite{Song:2023a}, with \tc s of 126~K at 50~GPa and 110~K at 80~GPa, respectively. Both of these phases can be derived from the $Fm\bar{3}m$ LaH$_{10}$ superhydride by removing a hydrogen atom from the 8c Wyckoff site and inserting a B/Be atom into the lattice so as to minimize the internal chemical pressure~\cite{Zurek:2022f}. 

A third and final strategy, which forms the basis of our study, involves predicting and synthesizing ternary hydrides that assume structures resembling those illustrated in Figure~\ref{fig:strucs}, but containing more than one type of  metal atom. This approach has many advantages. First, by choosing elements with different valences it may be possible to tune the position of the Fermi level, $E_\text{F}$, so that it sits on a peak in the density of states (DOS), thereby increasing the number of electrons that can participate in the superconducting mechanism, yielding a higher \tc . For example, a high-throughput study has predicted superconductivity in a number of $Pm\bar{3}m$ MXH$_{12}$ high-pressure hydrides with CaLuH$_{12}$ (\tc\ = 239~K~@ 160~GPa), YLuH$_{12}$ (\tc\ = 247~K~@ 160~GPa), and ScLuH$_{12}$ (\tc\ = 287~K~@ 190~GPa) characterized by van Hove singularities near $E_\text{F}$~\cite{Fam:2025a}. Because Y and Sc possess one additional valence electron than Ca, the latter two compounds have a higher DOS at $E_\text{F}$ than the former, in-line with their increased \tc s. A second potential advantage of this class of compounds is a lack of metal atom site preference, which increases the configurational entropy, $S_{\text{config.}}$, favoring the formation of metal alloy superhydrides. Computations have investigated ordered multi-component hydrides derived from the parent lattices illustrated in Figure~\ref{fig:strucs}, such as [Sc$_x$Ca$_{(1-x)}$]H$_n$ ($n=4,6$)~\cite{Shi:2021a,Yuan:2024a}, [Y$_x$Ca$_{(1-x)}$]H$_n$ ($n=4,6,10$)~\cite{Zhao:2025a,Zhao:2022a,Liang:2019a,Xie:2019a,Ferreira:2024a}, [Y$_{x}$Sc$_{(1-x)}$]H$_n$ ($n=4,6$)~\cite{Shi:2024a,Sukmas:2022a},  [Y$_{x}$Zr$_{(1-x)}$]H$_n$ ($n=4,6,9$)~\cite{Zhao:2023a}, [Y$_{x}$TH$_{(1-x)}$]H$_n$ ($n=6,9$)~\cite{Ghaffar:2024a,ChenW:2024a}, [La$_{x}$Th$_{(1-x)}$]H$_n$ ($n=6,9,10$)~\cite{Song:2024a,ChenW:2024a} and [La$_{x}$Y$_{(1-x)}$]H$_n$ ($n=4,6,10$)~\cite{Semenok:2021a,Wu:2023a,Song:2021a}. Most have only considered $x=0.5$ compositions~\cite{Shi:2021a,Yuan:2024a,Zhao:2022a,Liang:2019a,Xie:2019a,Shi:2024a,Sukmas:2022a,Fam:2025a,Ghaffar:2024a}, though some studies on larger cells with $x=0.25,0.75$ have been reported~\cite{Zhao:2022a,Du:2022a,Zhao:2023a,Wu:2023a,ChenW:2024a,Song:2024a}. Notably, only a handful of these investigations have considered $S_{\text{config.}}$ in assessing thermodynamic stability~\cite{Song:2024a,Ferreira:2024a,Ghaffar:2024a,Semenok:2021a}, and a few studies have incorporated it to thermodynamically stabilize multi-component hydrides with more than two metal atoms~\cite{Ma:2025b,Zhang:2025a}.

If the metal atoms in the aforementioned systems are disordered, $S_{\text{config.}}$ will promote their thermodynamic stability~\cite{Toher:2019a}, facilitating the synthesis of medium- and high-entropy superhydrides, and enhancing superconducting properties. This approach has been used to access  a variety of superconducting multi-component hydrides at high pressure~\cite{Zhao:2023b}. For example, (La,Y)H$_{6}$ and (La,Y)H$_{10}$ were synthesized  by laser heating a La-Y alloy compressed to 170-196~GPa with ammonia borane~\cite{Semenok:2021a}. Notably, two of the parent compounds, YH$_{6}$~\cite{Wang:2022a,Troyan:2021a,Kong:2021a} and LaH$_{10}$~\cite{Drozdov:2019a,Zulu:2019a}, have been synthesized, while two more, LaH$_{6}$ and YH$_{10}$, have not been made and are computed to lie above the convex hull at 150-200~GPa~\cite{Liu:2017a}. The measured \tc\ of (La,Y)H$_{10}$ (253~K~@ 183~GPa and 244~K~@ 153~GPa)~\cite{Semenok:2021a,Marathamkottil:2025a} is comparable to that of LaH$_{10}$ (260~K~@ 180-200~GPa and 250~K~@ 170~GPa)~\cite{Drozdov:2019a,Zulu:2019a}. Furthermore, (La,Y)H$_{4}$~\cite{Bi:2022a} has been synthesized at 110~GPa with a \tc\ of 92~K. Not only can this compound be made at a lower pressure than YH$_{4}$~\cite{Shao:2021a}, but its \tc\ is higher and superconductivity within it can persist down to 80~GPa. Similarly, (La,Ce)H$_{9}$ (\tc\ = 148-178~K~@ 97-172~GPa)~\cite{Ma:2022c}, (La,Ce)H$_{9-10}$ (\tc\ = 176~K~@ 100~GPa)~\cite{Chen:2023a}, and Y$_{0.5}$Ce$_{0.5}$H$_{9}$ (\tc\ = 141~K~@ 130~GPa)~\cite{Chen:2024a} have \tc s that surpass those of CeH$_{9}$ and CeH$_{10}$~\cite{Chen:2021a}. The synthesis of medium-entropy alloy superhydrides such as (La,Y,Ce)H$_{10}$, with a maximum  \tc\ of 190~K~\cite{Chen:2024b} and 165~K~\cite{Ma:2025a}, has been reported, paving the road towards the realization of high-entropy high-pressure hydride superconductors.

Herein, we present first-principles calculations on the (Y,Ca)H$_n$ system, which has previously been studied theoretically~\cite{Zhao:2025a,Zhao:2022a,Liang:2019a,Xie:2019a}, though $S_{\text{config.}}$ was only considered for the $n=6$ system~\cite{Ferreira:2024a}. We report the results of extensive evolutionary algorithm (EA) searches performed between 100-300~GPa for a wide hydrogen content and an equimolar composition of the two metal atoms. Our results illustrate how the identity of the metal atoms can be used to control doping of the hydrogenic lattice, and concomitantly the resulting \tc , and highlighting the importance of $S_{\text{config.}}$ on phase stability for some stoichiometries. For thermodynamically stable systems, and metastable phases that lay within 10~meV/atom of the convex hull, a semi-empirical RS5~\cite{Belli:2025a} fit was employed to estimate their {\tc}s, followed by electron-phonon coupling calculations on structures with potentially high {\tc}s to confirm their superconducting properties.

\section{Computational Details}
CSP searches were performed using the open-source \textsc{XtalOpt}~\cite{Zurek:2014d,Lonie:2011a,Avery:2019a,Falls:2021a,Hajinazar:2024a} EA v12.1 and v13 to uncover stable and metastable phases. Evolutionary runs were performed for the following stoichiometries and formula units (FUs) at 100, 150, 200, 250, and 300~GPa: YCaH$_{n}$ ($n = 8-20$) with 1-4 FUs and YCa with 1-8 FUs. Seeds were employed in EA searches for: (i) YCa  alloys (seeded with elemental phases where some of the Ca/Y atoms were substituted to obtain the desired stoichiometry), and (ii) YCaH$_{20}$ (seeded with $P4/mmm$ YCaH$_{20}$, derived from Ca-substituted $Fm\bar{3}m$ YH$_{10}$).  The initial generation of random symmetric structures was created using the RandSPG algorithm.~\cite{Avery:2017a} The minimum interatomic distances were constrained via summing atomic radii calculated by using a uniform scaling factor of 0.5 multiplied by tabulated covalent radii, with an absolute minimum radius of 0.25~\r{A}. Duplicate structures were identified via the \textsc{XtalComp}~\cite{Lonie:2012a} algorithm, and were eliminated from the breeding pool. The employed breeding operations and their relative probabilities were: stripple (50\%), permustrain (35\%), and crossover (15\%). The maximum number of generated structures was set to 1000 per run.

Geometries were relaxed using density functional theory (DFT) by the Vienna \textit{Ab-Initio} Simulation Package (VASP) v5.4.1 and v6.4.2~\cite{Kresse:1993a,Kresse:1996a,Kresse:1996b} with the gradient-corrected exchange and correlation functional of Perdew–Burke–Ernzerhof (PBE)~\cite{Perdew:1996a} and the projector augmented wave method~\cite{Blochl:1994a,Kresse:1999a}. The Ca~3p\textsuperscript{6}4s\textsuperscript{2}, H~1s\textsuperscript{1}, and Y~4s\textsuperscript{2}4p\textsuperscript{6}5s\textsuperscript{2}4d\textsuperscript{1} electrons were treated explicitly in all of the calculations. Each optimization consisted of four subsequent steps of increasing precision where the plane-wave  basis set energy cutoff varied from 350-400~eV and the $\pmb{k}$-point grid, generated using the $\Gamma$-centered Monkhorst-Pack scheme~\cite{Monkhorst:1976a}, varied so that the number of divisions along each reciprocal lattice vector was selected so that the product of this number with the real lattice constant was 14-30~{\AA}. Unique structures that were within 150~meV/atom of the lowest-enthalpy system in each completed EA search were further analyzed, as detailed below.

Precise calculations, including geometry optimizations, electronic band structures, DOS, electron localization functions (ELFs)~\cite{Becke:1990a,Silvi:1994}, and Bader charges were performed with a plane-wave basis set energy cutoff of 700~eV. A $\Gamma$-centered Monkhorst-Pack $\pmb{k}$-mesh was employed~\cite{Monkhorst:1976a}, and the number of divisions along each reciprocal lattice vector was selected so that the product of this number with the real lattice constant was 50~{\AA} for geometry optimizations and 70~{\AA} for electronic structure calculations. 
Bader charge analysis was performed using the Bader program v1.04.~\cite{Henkelman:2006a,Sanville:2007a,Tang:2009a}

The dynamic stability of the predicted structures was determined via the supercell approach~\cite{Parlinski:1997a,Chaput:2011a} using the Phonopy open-source package~\cite{Togo:2015a,Togo:2023a} interfaced with VASP. The supercell sizes employed contained between 88-384 atoms for the YCaH$_n$ phases, with lattice vector lengths of at least 10~\AA{} (when computationally feasible). Finite temperature effects renormalize unit cell parameters. This phenomenon can be modelled using the quasi-harmonic approximation; however, because inclusion of volume-dependent thermal effects requires numerous phonon calculations, it was deemed to be computationally too expensive for this study, where numerous phases and pressures were considered. Imaginary frequencies, which likely resulted from interpolation errors, were excluded from the thermodynamic property calculations used to obtain the zero-point energy (ZPE) and vibrational entropy ($S_{\text{vib.}}$). Additionally, the thermodynamic property calculations neglected the contribution of the electronic entropy.

The {\tc} was predicted via TcESTIME~\cite{Novoa:2025a} with the RS5 fit~\cite{Belli:2025a} using the electronic DOS from VASP calculations and Critic2~\cite{Otero-de-la-Roza:2009a,Otero-de-la-Roza:2014a} to compute the ELF critical points. The electron-phonon coupling (EPC) and density functional perturbation theory (DFPT) calculations were performed using the Quantum ESPRESSO (QE) program~\cite{Giannozzi:2009a,Giannozzi:2017a,Giannozzi:2020} v7.0, v7.3, and v7.4.1. The interpolation scheme~\cite{Wierzbowska:2006a}, where the EPC matrix elements were computed using a small $\pmb{k}$-point grid and interpolated linearly to a denser grid for each $\pmb{q}$-point, was employed. The {Ca 3s\textsuperscript{2}3p\textsuperscript{6}4s\textsuperscript{2}}, {H 1s\textsuperscript{1}}, and {Y 4s\textsuperscript{2}4p\textsuperscript{6}5d\textsuperscript{2}4d\textsuperscript{1}} ultrasoft pseudopotentials were obtained from the PSlibrary v1.0.0, and generated using the `atomic' code v7.0 by Dal Corso.~\cite{Dalcorso:2014a} The plane-wave basis set energy cutoff was chosen to be 80~Ry. The Brillouin zone sampling scheme of Methfessel-Paxton\cite{Methfessel:1989a} was used. Additional details on the chosen computational parameters of the EPC calculations are provided in the Supplementary Information. The \tc\/s were calculated via the Allen-Dynes modified McMillan equation~\cite{Allen:1975a} with strong-coupling and shape corrections, and by numerically solving the isotropic Eliashberg equations~\cite{Eliashberg:1960a}, as implemented in the IsoME~\cite{IsoME} software package v1.0.4 with the constant DOS approximation, and renormalized Coulomb potentials, $\mu^{*}$, of 0.10 and 0.13.

\section{Results and Discussion}
\subsection{Thermodynamic Stability and the Cold Phase Diagram}
\begin{figure}[!htb]
    \centering
    \includegraphics[width=17 cm]{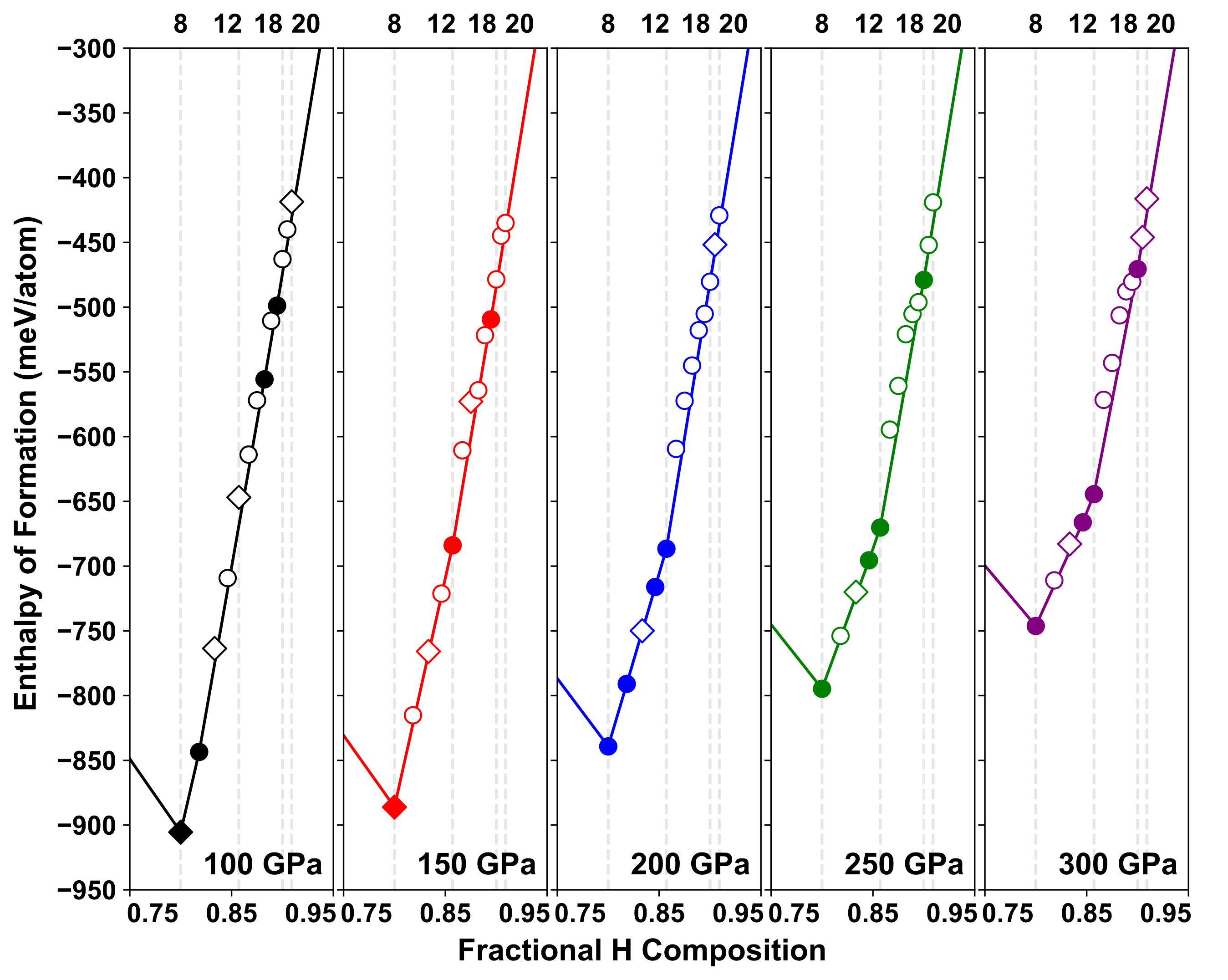}
\caption{${\Delta}H_{\text{F}}$ for the reaction $\left( \frac{1}{2}\text{H}_{2} \right)_{n} + \text{Ca} + \text{Y} \to \text{YCaH}_{n}$ versus the mole fraction of $\frac{1}{2}$H$_{2}$ in the ternary as a function of pressure from 100-300~GPa, calculated using the enthalpies of the $P6_{3}/m$ (100~GPa)~\cite{Pickard:2007a}, $C2/c$ (150-250~GPa)~\cite{Pickard:2007a,Gorai:2023a}, and $Cmca$ (300~GPa)~\cite{Pickard:2007a,Gorai:2023a} phases for H$_{2}$; $P4_{3}2_{1}2$ (100~GPa)~\cite{Ishikawa:2010a}, $Cmce$ (150-250~GPa)~\cite{Ishikawa:2010a}, and $I4/mcm$ (300~GPa)~\cite{Ishikawa:2010a,Arapan:2008a} phases for Ca; and the $Fddd$~\cite{Li:2019a,Buhot:2020a,Pace:2020a} phase for Y. Thermodynamically stable (unstable) structures are displayed by solid (hollow) shapes. Structures that are dynamically stable (unstable) in the harmonic approximation are shown as circles (diamonds).}\label{fig:eform_enthalpy_all}
\end{figure}
Synthesized binary high-pressure hydrides include compounds predicted by DFT to lie on the zero-Kelvin convex hull (e.g.\ YH$_{4}$~\cite{Shao:2021a,Troyan:2021a,Wang:2022a}, YH$_{6}$~\cite{Troyan:2021a,Kong:2021a,Wang:2022a}, YH$_{9}$~\cite{Kong:2021a,Wang:2022a}, CaH$_{4}$~\cite{Mishra:2018a}, and CaH$_{6}$~\cite{Ma:2022b,Li:CaH6}), as well as DFT-predicted metastable compounds (e.g.\ $C2/m$ Ca$_{2}$H$_{5}$~\cite{Mishra:2018a}, and  $Imm2$ YH$_{7}$~\cite{Troyan:2021a,Wang:2022a}). We wondered if these structure-types, or completely new ones, would emerge as being stable or metastable in our ternary hydride EA searches? To answer this question we began by plotting the calculated enthalpies of formation, ${\Delta}H_{\text{F}}$, of the most favorable YCaH$_{n}$ phases found at 100-300~GPa in a 2D representation (Figure~\ref{fig:eform_enthalpy_all}), within the static lattice approximation. At 100~GPa YCaH$_{8}$, YCaH$_{9}$, YCaH$_{15}$, and YCaH$_{17}$ were predicted to be thermodynamically stable. At 150~GPa, YCaH$_{8}$ and YCaH$_{17}$ remained on the hull, while YCaH$_{12}$ replaced YCaH$_{15}$. Between 200-300~GPa, both YCaH$_{8}$ and YCaH$_{12}$ were still on the hull, where they were joined by YCaH$_{11}$. YCaH$_{9}$ (200~GPa) and YCaH$_{18}$ (250-300~GPa) were also found to be thermodynamically stable at the pressures given in the braces. Notably, all of the points included in Figure~\ref{fig:eform_enthalpy_all} were within 40~meV/atom of the hull, an amount that could be overcome by a different choice of exchange-correlation functional, or the inclusion of zero-point, anharmonic, or finite temperature effects.

Phonon calculations, within the harmonic approximation, were employed to confirm the dynamic stability (Figures~\ref{fig:YCaH8_Cmmm_phonons}-\ref{fig:YCaH20_P4.nmm_phonons}) of low-enthalpy structures uncovered from the EA searches (and some that were manually constructed from plausible structure types), and the ones verified as being local minima were employed to generate ternary convex hulls excluding (Figures~\ref{fig:YCaHn_TernaryConvexHullEnthalpy_100GPa}-\ref{fig:YCaHn_TernaryConvexHullEnthalpy_300GPa}) and including (Figures~\ref{fig:YCaH_convex_zoom_vib_100GPa}(b)-\ref{fig:YCaH_convex_zoom_vib_300GPa}(b)) zero-point effects. In both cases, the enthalpies of a majority of the dynamically stable phases were found to lie within 67~meV/atom of the 3D convex hulls, falling within the 90$^\text{th}$ percentile of metastability for inorganic crystalline materials at zero pressure and 0~K~\cite{materialsproject}.
In what follows, we consider the role of the ZPE and temperature, calculated within the harmonic approximation, on stability. 
For YCaH$_{n}$ stoichiometries with $n = 8, 12, 18, 20$ multiple structures whose enthalpies differed by only 1-15~meV/atom were found at certain pressures (Figures~\ref{fig:relative_enthalpies}a, \ref{fig:relative_enthalpies}e, \ref{fig:relative_enthalpies}k, and \ref{fig:relative_enthalpies}m). Their hydrogenic lattices adopted one of the structure types illustrated in Figure~\ref{fig:strucs}, and the difference between them stemmed only from the way in which the metal lattice was ``colored,'' hinting that $S_{\text{config.}}$ may be important for this ternary system, as analyzed below.

\subsection{Two Distinct YCaH$_8$ Structure Types}
\begin{figure}[!htb]
    \centering
    \includegraphics[width=0.75\columnwidth]{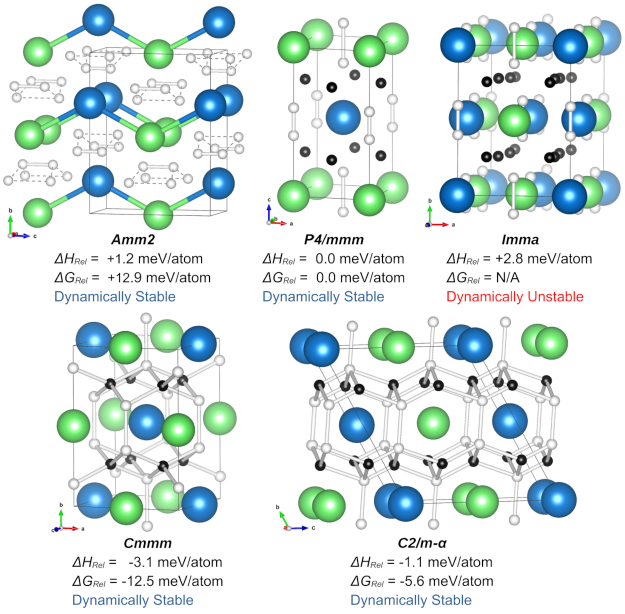}
\caption{The $Amm2$, $P4/mmm$, $Imma$, $Cmmm$ and $C2/m$-$\alpha$ YCaH$_{8}$ structures (Ca/Y atoms are blue/green and apical/basal H atoms are white/black) at 200~GPa.  $Amm2$ is derived from $Cmcm$ MH$_{4}$, while the others are derived from $I4/mmm$ MH$_4$ (Figure~\ref{fig:strucs}). 
The dynamic stability at this pressure, calculated within the harmonic approximation, is provided, as are the relative enthalpies and Gibbs free energies (including ZPE and $S_{\text{vib.}}$ at 1500~K) with respect to $P4/mmm$.}\label{fig:YCaH8_energy_compare}
\end{figure}

The ordered YCaH$_{8}$ phases found in our EA searches, derived from the two parent MH$_{4}$ lattices, are illustrated in Figure~\ref{fig:YCaH8_energy_compare}. At 200~GPa their enthalpies were within 6~meV/atom of each other, and at 100~GPa and 300~GPa, this enthalpy window increased to 11~meV/atom and 15~meV/atom, respectively. The $Amm2$ phase is based off of the $Cmcm$ MH$_4$ structure (Figure~\ref{fig:strucs}(b)), which is predicted to be the most stable phase for CaH$_{4}$ above 180~GPa~\cite{Wang:2012a} and SrH$_4$ above 85~GPa~\cite{Hooper:2014}. $Amm2$ YCaH$_8$ becomes competitive (Figure~\ref{fig:relative_enthalpies}(a)) with the (remaining) $I4/mmm$ based YCaH$_8$ phases around 180-300~GPa. Specifically, above 180~GPa the enthalpy of $Amm2$  is within 8~meV/atom of the most stable YCaH$_8$ structure, while at 160~GPa this difference increases to $\sim$17~meV/atom. Although no other $Cmcm$-based phases were found in our EA searches, we manually constructed two with $C2/m$ and $P2_{1}/m$ symmetries (Figure~\ref{fig:YCaH8_phases_from_Cmcm_MH4}). After a geometry optimization at 200~GPa, they were within 6~meV/atom of $Amm2$, and at 100~GPa, they were within 5~meV/atom of it.

The enthalpies of the other four phases, derived from various colorings of $I4/mmm$ MH$_4$, which was synthesized and predicted for many single-element tetrahydrides~\cite{Bi:2021a}, were within 20~meV/atom of one another between 100-300~GPa, hinting of potential disorder on the metal atom sites. In $P4/mmm$ YCaH$_{8}$ the Y and Ca atoms lie on distinct $ab$ planes, whereas in the other $I4/mmm$-type structures each plane possesses an equal fraction of the two types of metals. As a result, in the $P4/mmm$ phase the hydrogen atoms that lie on the apical (H$_a$) positions form H$_2$ molecules with two distinct bond lengths: those that are oriented side-on to the Y atoms are elongated as compared to those that surround the Ca atoms  (1.36 vs.\ 0.94~\AA{} at 200~GPa).  The longer H-H distance is likely caused by the increased charge donation from yttrium (which can formally adopt a +3 oxidation state) than from calcium (with a formal +2 oxidation state), to the nearby H$_2$ molecules, thereby partially filling dihydrogen's antibonding $\sigma^*$-orbitals. This difference in charge donation makes YH$_4$ a good metal with a high \tc , while CaH$_4$ is not~\cite{Bi:2021a}.  The H$^-$ units are located on the so-called basal sites, H$_b$, and the distance between them and the H$_a$ hydrogens that are side-on to Y (Ca) measures 1.47~\AA{} (1.53~\AA{}) at 200~GPa. 
In the $Imma$ phase, each of the H$_a$ atoms also comprise H$_2$ units, but here they are oriented side on to three metal atoms of one type and one metal atom of the other type, resulting in two distinct bond distances of 1.19~\AA{} and 0.98~\AA{} at 200~GPa.  The distances between the H$_{b}$ atoms varies from 1.67 to 2.13~\AA{}, while the H$_{a}$-H$_{b}$ distances fall between 1.45-1.59~\AA{} at this pressure.

$Cmmm$ and $C2/m$-$\alpha$ YCaH$_8$ are distinctly different than the aforementioned configurations since the contacts between the apical hydrogen atoms are not the shortest above $\sim$170-180~GPa. The basal and apical hydrogens within $Cmmm$ form puckered 1D chains with H$_a$-H$_b$ distances of 1.38 and 1.45~\AA{}, which are joined to the neighboring layer of puckered chains by H$_a$-H$_a$ contacts measuring 1.47 and 1.55~\AA{} at 200~GPa.  Another way to describe this structure would be as a clathrate with four distorted hexagonal faces whose sides measure 1.38-1.55~\AA{}, capped on either side by two rhombi (whose sides measure 1.38~\AA{}) and two kites (with 1.38 and 1.45~\AA{} sides). The $C2/m$-$\alpha$ structure is very similar to $Cmmm$ YCaH$_8$, except it is less symmetrical, so that the edges of both the hexagonal faces comprising the clathrate and the quadrilaterals have a larger spread, falling between 1.14-1.56~\AA{} at 200~GPa.

Within the harmonic approximation and neglecting (likely) phonon interpolation errors, all of these phases were found to be dynamically stable at 200~GPa (Figures~\ref{fig:YCaH8_Cmmm_phonons}-\ref{fig:YCaH8_Amm2_phonons}) -- except for $Imma$, which becomes stabilized at 210~GPa -- and they remain dynamically stable up to 300~GPa. Both $Cmmm$ and $C2/m$-$\alpha$ are dynamically unstable at 150~GPa, while at 100~GPa only the $Amm2$ phase maintained dynamic stability. Though including the ZPEs calculated within the harmonic approximation did affect the relative enthalpies of these phases somewhat (Table~\ref{tab:YCaH_ConvexHullDistances_200GPa}), the differences between them were still small, suggesting they may exist as solid solutions.  It has been pointed out that a regular solid solution of high-pressure metal hydrides would possess a random arrangement of solutes with the crystal symmetry of the parent phase, and that interactions between adjacent clusters would renormalize the phonons and locally stabilize any imaginary modes~\cite{Ferreira:2024a}. Neglecting such effects, Figure~\ref{fig:YCaH8_energy_compare} provides the relative Gibbs free energies including $S_{\text{vib.}}$, at 200~GPa and 1500~K (a temperature typically attained during laser heating), of the dynamically stable YCaH$_8$ phases. These contributions stabilize the $I4/mmm$ over the $Cmcm$ based geometries at these conditions, hinting that they may be preferentially synthesized, though the formation of mixed-phases in laser-heated diamond anvil cells is common.

\subsection{Solid Solutions of (Y,Ca)H$_6$ in the Sodalite Structure}
\begin{figure}[!htb]
    \centering
    \includegraphics[width=0.7\columnwidth]{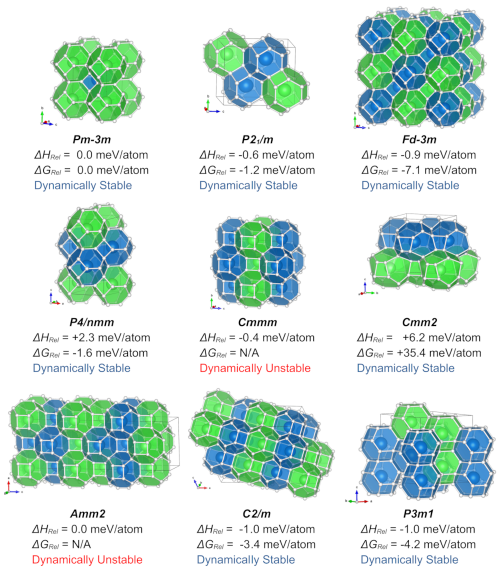}
\caption{Various ordered YCaH$_{12}$ structures at 150~GPa, with space groups noted. Ca/Y/H atoms are blue/green/white. Connections drawn between hydrogen atoms highlight the clathrate structure. The dynamic stability at this pressure, calculated within the harmonic approximation, is provided, as are the enthalpies and Gibbs free energies (including ZPE and $S_{\text{vib.}}$ at 1500~K) with respect to $Pm\bar{3}m$.}
\label{fig:YCaH12_energy_compare}
\end{figure}

The sodalite-like CaH$_6$ phase was the first high-pressure high-\tc\ clathrate to be predicted~\cite{Wang:2012a}, but it took more than a decade before its synthesis was reported~\cite{Ma:2022b,Li:CaH6}. Since then, numerous ordered ternary MH$_6$-based structures have been predicted computationally including $Pm\bar{3}m$, $P2_{1}/m$, $Fd\bar{3}m$, $P4/nmm$, and $Cmmm$ symmetry YCaH$_{12}$ phases~\cite{Zhao:2025a,Liang:2019a,Xie:2019a}, and it was noted that their enthalpies were quite competitive at megabar pressures~\cite{Liang:2019a}. In addition to these structures, our evolutionary searches found three (or four)  more  (Figure~\ref{fig:YCaH12_energy_compare}), which could be understood as different binary colorings of $Im\bar{3}m$ MH$_{6}$ (Figure~\ref{fig:strucs}(c)).  At 100~GPa, the enthalpies of all of these YCaH$_{12}$ geometries were within 9~meV/atom, and this difference decreased to $<$~6~meV/atom at 300~GPa (Figure~\ref{fig:relative_enthalpies}(e)). At 150~GPa, seven of these phases were nearly isoenthalpic, within 1~meV/atom of each other.  $Fd\bar{3}m$ YCaH$_{12}$  was found to have the lowest enthalpy between 160-300~GPa, in-line with previous calculations identifying it as the thermodynamic ground state between 125-400~GPa~\cite{Zhao:2025a,Liang:2019a}. Nonetheless, the small enthalpic differences between ordered variants hint of the possibility of disorder on the metal atom sites.

The atomic radius of Ca is estimated to be somewhat smaller than that of Y under pressure (1.44 and 1.28~\AA{} vs.\ 1.53 and 1.34~\AA{} at 100 and 300~GPa, respectively)~\cite{Rahm:2020a}. In the binary MH$_6$ systems, the clathrate cages possess a single distinct H-H distance (Figure~\ref{fig:strucs}(c)) measuring 1.24~\AA{} in CaH$_6$ and 1.27~\AA{} in YH$_6$ at 150~GPa, in-line with expectations based on their relative radii. In the simplest YCaH$_{12}$ structure, $Pm\bar{3}m$, the H-H distances of the square faces comprising the polyhedron that surround the Ca (1.23~\AA{}) and Y (1.29~\AA{}) atoms is quite similar to the values found in the binary MH$_6$ systems.  In the remaining ordered  YCaH$_{12}$ phases the arrangement of the metal atoms results in larger deviations of the bond lengths (e.g., with H-H distances measuring between 1.10-1.31~\AA{}), and distortions of the truncated octahedra from perfect symmetry results in faces of different quadrilateral  and  hexagonal shapes.

$Cmm2$ YCaH$_{12}$, which has the lowest enthalpy below 120~GPa (Figure~\ref{fig:relative_enthalpies}(e)), is noticeably distorted compared to the other YCaH$_{12}$ phases. In fact, when this structure underwent a geometry optimization above 150~GPa, it transformed into $Cmmm$ YCaH$_{12}$, but at 100~GPa, $Cmm2$ was found to be dynamically unstable within the harmonic approximation (Figure~\ref{fig:YCaH12_Cmm2_phonons}(a)). We hypothesize that this complicated potential energy surface results from our neglect of anharmonicity and quantum nuclear effects, which, if considered, would yield only the highly symmetric $Cmmm$ minimum. A similar situation was found for $Fm\bar{3}m$ LaH$_{10}$ where DFT calculations predicted distorted structures would be preferred below 230~GPa, but the inclusion of quantum effects was key for the stabilization of the symmetric phase to lower pressures, in agreement with experiment~\cite{errea2020quantum}. Furthermore, our calculations showed that all ordered YCaH$_{12}$ phases are dynamically unstable at 100~GPa (Figure~\ref{fig:YCaH12_Cmmm_phonons}-\ref{fig:YCaH12_P3m1_phonons}), while at 150~GPa, only two phases ($Cmmm$ and $Amm2$) do not correspond to local minima. With the exception of $Cmm2$, all of the phases considered were dynamically stable between 200-300~GPa.

As the laser heating conditions used to synthesize these phases can reach temperatures in the thousands-of-Kelvins, the relative Gibbs free energies were computed incorporating the ZPE and $S_{\text{vib.}}$ obtained within the harmonic approximation. Though the $Fd\bar{3}m$ 
phase is the most stable configuration at 150~GPa and 1500~K, the difference between it and $Pm\bar{3}m$ YCaH$_{12}$ is only $\sim$7~meV/atom, a value that is significantly smaller than the Boltzmann energy ($k_\text{B}T$) at this temperature, once again suggesting the possibility of solid solution behavior on the metal atom sites.

\subsection{Ordered or Disordered YCaH$_{18}$ and YCaH$_{20}$ Phases?}
The $P6_3/mmc$ YH$_9$~\cite{Du:2023a,Peng:2017a} and $Fm\bar{3}m$ YH$_{10}$~\cite{Liu:2017a} phases (and distorted variants thereof~\cite{Du:2023a}) have been predicted to be (meta)stable under pressure, and synthesis of the nonahydride has been reported~\cite{Kong:2021a}. The same is not true of their Ca-containing analogues, where the only clathrate-like phase predicted to lie on or near the convex hull, which was also synthesized, is $Im\bar{3}m$ CaH$_6$~\cite{Wang:2012a,Ma:2022b,Li:CaH6}. The hydrogenic lattice of the most stable CaH$_9$ phase found, which lies on the convex hull between 235-400~GPa and possesses the $C2/m$ space group, has not been described as a clathrate~\cite{Shao:2019a}. 
Instead, its hydrogenic lattice, which can be derived from a distortion of $P6_3/mmc$ CaH$_9$~\cite{Shao:2019a}, contains hydridic hydrogens, H$_2$ molecules, and bent H$_3$ units. The lowest enthalpy CaH$_{10}$ phase is metastable, possessing $R\bar{3}m$ symmetry, and its structure can be described as Ca atoms sandwiched between puckered graphene-like H-layers~\cite{Shao:2019a}, similar to a previously predicted SrH$_{10}$ structure~\cite{Wang:2015a}. The stability of clathrate lattices in YH$_9$ and YH$_{10}$, and preference for the formation of lower dimensional hydrogenic lattices in CaH$_9$ and CaH$_{10}$, was rationalized in terms of the formal effectively added electron concept~\cite{Wang:2012a}, by noting that yttrium donates a larger amount of charge to the hydrogenic lattice than calcium~\cite{Shao:2019a}. This situation is therefore different than what was found for the $I4/mmm$ MH$_4$ and $Im\bar{3}m$ MH$_6$ structures types, which are stable for both calcium and yttrium, within their respective pressure ranges. We therefore wondered if (disordered) clathrate-like YCaH$_{18}$ and YCaH$_{20}$ phases might be stable under pressure, or if the internal chemical pressures~\cite{Zurek:2022f} would prevent such compounds from forming?

\begin{figure}[!htb]
    \centering
    \includegraphics[width=0.6\columnwidth]{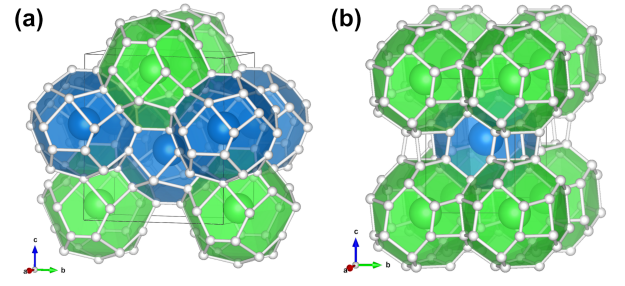}
\caption{The ordered (a) $Pmmn$-$\alpha$ YCaH$_{18}$ and (b) $P2$ YCaH$_{20}$ phases at 200 and 300~GPa, respectively. Ca/Y/H atoms are blue/green/white. Connections drawn between hydrogen atoms highlight the clathrate structure.}
\label{fig:dynamically_stable_YCaH18_YCaH20}
\end{figure}

Among the YCaH$_{18}$ structures that were uncovered, only the $Pmmn$-$\alpha$ phase illustrated in Figure~\ref{fig:dynamically_stable_YCaH18_YCaH20}(a) was found to be dynamically stable between 200-300~GPa (Figure~\ref{fig:YCaH18_Pmmn-alpha_phonons}), and it was the lowest-enthalpy structure above 170~GPa (Figure~\ref{fig:relative_enthalpies}(k)). Though the other phases emerging in our evolutionary searches were dynamically unstable within the harmonic approximation, we briefly mention them, because of the close analogy with the previously proposed distorted variants of $P6_3/mmc$ YH$_9$~\cite{Du:2023a}, and because anharmonicity could impact these distortions. $Pmmn$-$\beta$ YCaH$_{18}$ (Figure~\ref{fig:YCaH18_energy_compare}) was found to be structurally similar to its $\alpha$ variant, but it was dynamically unstable between 200-300~GPa  (Figure~\ref{fig:YCaH18_Pmmn-beta_phonons}). From 100-300~GPa an $Amm2$ symmetry phase (Figure~\ref{fig:YCaH18_energy_compare}) with a different ordering of metal atoms was predicted to lie within 4~meV/atom of $Pmmn$-$\alpha$, but within the harmonic approximation its dynamic stability could not be confirmed. Another phase, $P\bar{6}m2$, resembled $Amm2$, but it was also dynamically unstable in the same pressure range (Figure~\ref{fig:YCaH18_P-6m2_phonons}). Additionally, we manually constructed two other structures, $P\bar{3}m1$ and $Pmma$ (Figure~\ref{fig:YCaH18_phases_from_hex_MH9}), and although they were enthalpically competitive to $Pmmn$-$\alpha$ at 200~GPa (within 10~meV/atom) and 300~GPa ($\sim$1~meV/atom), they were also found to be dynamically unstable between 200-300~GPa (Figures~\ref{fig:YCaH18_P-3m1_phonons} and \ref{fig:YCaH18_Pmma_phonons}). Though more detailed investigations that explicitly consider a wider variety of ordered YCaH$_{18}$ phases, and effects of anharmonicity, would be required for confirmation, our results support the stability of only a single ordered phase ($Pmmn$-$\alpha$ YCaH$_{18}$). Since our calculations do not suggest metal atom disorder, $S_{\text{config.}}$ will not be included in the construction of (ternary) convex hulls at finite temperatures for the YCaH$_{18}$ stoichiometry.

In the case of YCaH$_{20}$ the only phase found to be dynamically stable (at 300~GPa, but not at 250~GPa) was the $P2$ structure (Figure~\ref{fig:dynamically_stable_YCaH18_YCaH20}(b))---a highly distorted variant of the $P4/mmm$ symmetry system (Figure~\ref{fig:YCaH20_energy_compare}), which has been reported to be dynamically stable at 550~GPa~\cite{Zhao:2022a}. Because the parent $Fm\bar{3}m$ CaH$_{10}$ phase is not the most stable configuration up to 300~GPa, it is unclear if the inclusion of quantum nuclear effects, which are beyond the scope of this study, would symmetrize the $P2$ structure yielding the $P4/mmm$ at lower pressures. Two other colorings of the (undistorted) parent clathrate structure, with $I4_{1}/amd$ and $R\bar{3}m$ symmetry (Figure~\ref{fig:YCaH20_energy_compare}), were determined to be within 5~meV/atom of $P4/mmm$ from 150-300~GPa (Figure~\ref{fig:relative_enthalpies}(m)); however, their phonons indicated that they are dynamically unstable at 300~GPa (Figures~\ref{fig:YCaH20_I4_1.amd_phonons} and \ref{fig:YCaH20_R-3m_phonons}). Furthermore, other YCaH$_{20}$ structures (Figure~\ref{fig:YCaH20_phases_fcc_MH10}) were constructed manually -- $P\bar{3}m1$, $P3m1$, $P4_{2}/mmc$, $Pmm2$, and $P4/nmm$ -- though their enthalpies within 4~meV/atom of $P4/mmm$ at 300~GPa, they were determined to be dynamically unstable at this pressure (Figures~\ref{fig:YCaH20_P-3m1_phonons}-\ref{fig:YCaH20_P4.nmm_phonons}). These results suggest that, similar to YCaH$_{18}$, $S_{\text{config.}}$ should not be considered in the construction of ternary convex hulls that include YCaH$_{20}$, as only a single ordered structure was found to be dynamically stable.

\subsection{Computed Phase Diagrams}

In the past, thermodynamic stability for high-pressure hydrides has been deduced computationally by plotting their formation enthalpies, calculated at 0~K for static nuclei, and determining which phases comprise the convex hulls. Low-lying metastable systems have also often been considered in these analyses~\cite{Zurek:2019a}. Furthermore, it has been illustrated that the effect of the ZPE, which can be significant for light nuclei, can alter the identity of the thermodynamically stable high pressure hydrides~\cite{Ye:2018a,Wang:2012a,Liang:2019a,Zurek:2024f}. In our 0~K results, and neglecting the ZPE,  $Fd\bar{3}m$ YCaH$_{12}$ lies on the 3D convex hull between 150-300~GPa. At 150~GPa it is accompanied by a dynamically stable $Cm$-$\alpha$ YCaH$_{17}$ phase (Figure~\ref{fig:YCaH17_Cm-alpha_phonons}(b)) consisting of H$_{2}$ units and H$^{-}$ anions, but inclusion of the ZPE removes YCaH$_{17}$ from the hull. 
In fact, including ZPE generally increases the convex hull distances for most phases with low symmetries (Tables~\ref{tab:YCaH_ConvexHullDistances_100GPa}-\ref{tab:YCaH_ConvexHullDistances_300GPa}). Meanwhile, for stoichiometries that are susceptible to metal atom disorder, we find that the ZPE may either decrease their convex hull distance, like for YCaH$_{8}$, or have little to no effect on thermodynamic stability, as seen with YCaH$_{12}$.

At finite temperature, $S_{\text{vib.}}$ is a component of the free energy. As the number of species in a compound increases, the enthalpic stabilization often decreases, and $S_{\text{config.}}$ becomes progressively important, provided that (partial) disorder does not bear too-high of an energetic penalty~\cite{Toher:2019a}. The importance of $S_{\text{config.}}$ and $S_{\text{vib.}}$ on thermodynamics has been investigated for high-entropy ceramics~\cite{Esters:2021a}, but its role for multi-component hydrides with potential disorder on the metal atom sites has only recently become a focus of computational studies~\cite{Ferreira:2024a}. However, because multi-component hydrides, such as (La,Y)H$_{10}$, can now be synthesized by laser heating metal alloys together with ammonia borane to thousands of degrees Kelvin~\cite{Marathamkottil:2025a,Semenok:2021a,Bi:2022a,Ma:2022c,Chen:2023a,Chen:2024b,Ma:2025a}, it is important to study the role of $S_{\text{config.}}$ on their stability and properties.

The comparison of the enthalpies of YCaH$_n$ phases with different metal atom arrangements performed above suggests that YCaH$_8$ and YCaH$_{12}$ may have disordered metal lattices, while for YCaH$_{18}$ and YCaH$_{20}$, only a single ordered phase was viable. For the other stoichiometries, $S_{\text{config.}}$ was not considered as they were generally found to prefer a single ordered phase, or multiple enthalpically competitive phases that do not share the same hydrogen sublattice were found. Using these assumptions, we have calculated the free energies of the phases considered here as a function of $P$ and $T$ including the ZPE, $S_{\text{vib.}}$  (Figures~\ref{fig:YCaH_convex_zoom_vib_100GPa}-\ref{fig:YCaH_convex_zoom_vib_300GPa}, \ref{fig:YCaHn_FormationEnergy_vib_100GPa}-\ref{fig:YCaHn_FormationEnergy_Zoom_vib_300GPa}, and Tables~\ref{tab:YCaH_ConvexHullDistances_100GPa}-\ref{tab:YCaH_ConvexHullDistances_300GPa}), and $S_{\text{config.}}$, when applicable (Figures~\ref{fig:YCaH_convex_zoom_S-select_100GPa}-\ref{fig:YCaH_convex_zoom_S-select_300GPa}). In case of the latter, we employed the Boltzmann formula (simplified via Stirling's approximation) as $S_{\text{config.}}/k_\text{B}=\ln\frac{N!}{(N/2)!(N/2)!}=-\ln(1/2)$\cite{Rost:2015a}. This provides an upper bound to the stabilization that can be attained by randomly distributing the two metal atoms within the YCaH$_8$ and YCaH$_{12}$ clathrate cages. At 3500~K, $TS_{\text{config.}}$ was calculated as being $\sim$42 and $\sim$30~meV/atom for YCaH$_8$ and YCaH$_{12}$, respectively. 
In a real system with some site preference, $S_{\text{config.}}$ would be smaller; nonetheless, the value we calculate here can be used to investigate the importance on the $S_{\text{config.}}$ on phase stability.
\begin{figure}[!htb]
    \centering
    \includegraphics[width=1.0\columnwidth]{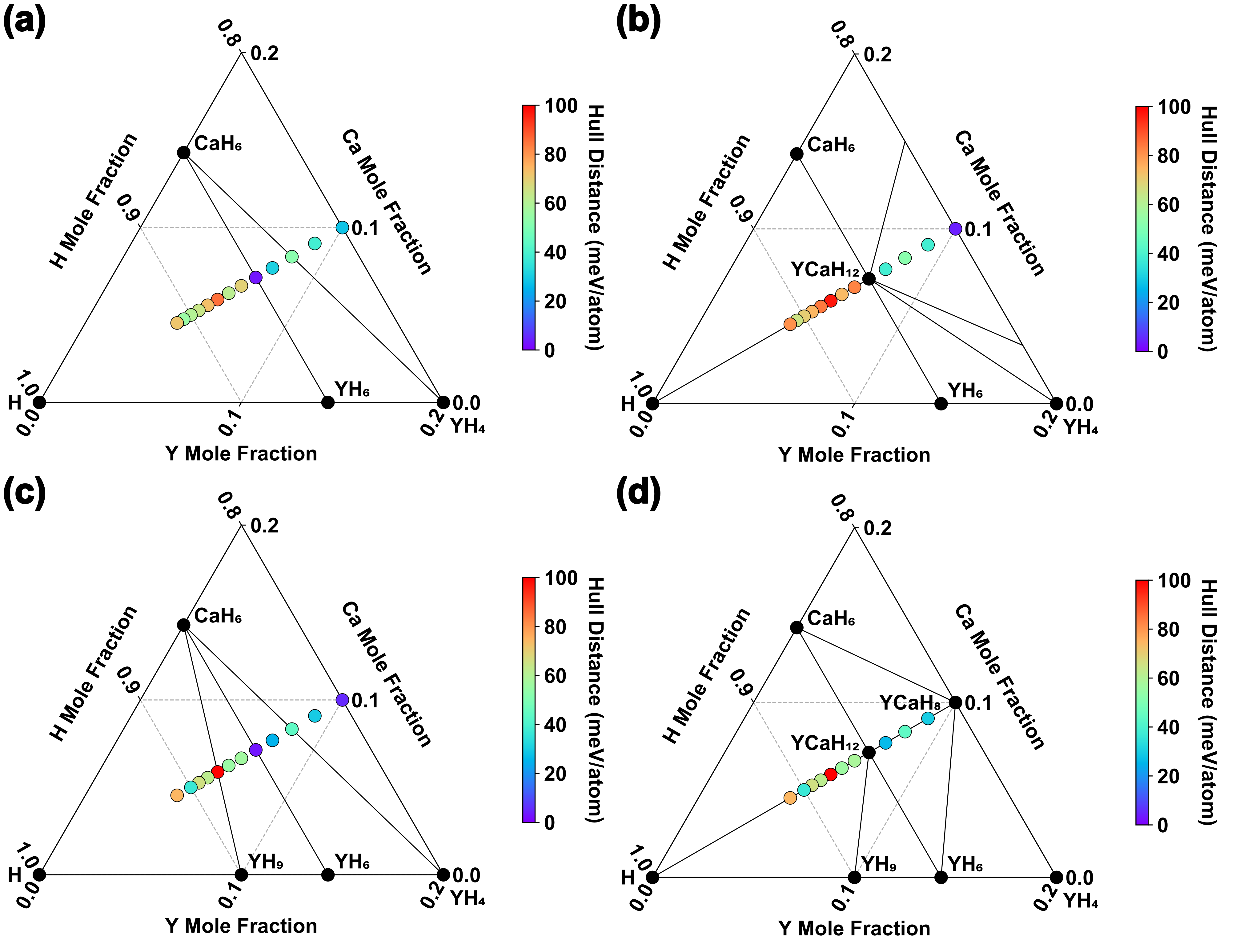}
\caption{The ternary (3D) convex hulls at 2000~K zoomed into regions with high H concentration at: (a) 150~GPa with ZPE + $S_{\text{vib.}}$ and with (b) $S_{\text{config.}}$; (c) 200~GPa with ZPE + $S_{\text{vib.}}$ and with (d) $S_{\text{config.}}$.}
\label{fig:zoomed_convex_hulls_comparison}
\end{figure}

At 150~GPa, inclusion of $S_{\text{vib.}}$ removes $Fd\bar{3}m$ YCaH$_{12}$ from the convex hull above 1500~K (Figure~\ref{fig:zoomed_convex_hulls_comparison}(a)); however, when $S_{\text{config.}}$ is also included, YCaH$_{12}$ lays on the hull between 2000-3500~K (Figure~\ref{fig:zoomed_convex_hulls_comparison}(b), Figure~\ref{fig:YCaH_convex_zoom_S-select_150GPa}(j)). 
Similarly, YCaH$_{12}$ no longer lies on the hull when ZPE and $S_{\text{vib.}}$ are included at 200~GPa and 2000~K (Figure~\ref{fig:zoomed_convex_hulls_comparison}(c)) and 250~GPa at 2500~K (Figure~\ref{fig:YCaH_convex_zoom_vib_250GPa}(h)). With $S_{\text{config.}}$, however, YCaH$_{12}$ is thermodynamically stable at 200~GPa and 2000~K (Figure~\ref{fig:zoomed_convex_hulls_comparison}(d)) and 250~GPa and 2500~K (Figure~\ref{fig:YCaH_convex_zoom_S-select_250GPa}(h)). YCaH$_{8}$ was also found to lay on the hull at 200~GPa and 2000~K (Figure~\ref{fig:zoomed_convex_hulls_comparison}(d)). 
At 300~GPa, YCaH$_{12}$ is predicted to be stable at all temperatures irrespective of the inclusion of $S_{\text{config.}}$, while YCaH$_8$ was found to lie within 0-26~meV/atom from the hull.

Below we summarize the observed trends from our calculations, where the ZPE and $S_{\text{vib.}}$ were estimated from the harmonic approximation using 0~K optimized structures. Inclusion of anharmonic and volume-dependent temperature effects may impact these results, however due to their computational expense, they lay outside the scope of our study. Since it is unknown how kinetics, local stoichiometric variations, and pressure anisotropies impact which phases are formed, low-lying metastable systems should be considered as potentially viable candidates. Briefly:
\begin{itemize}
\item At 100~GPa, no ternary hydride was predicted to be thermodynamically stable, though YCaH$_{15}$, YCaH$_{17}$, and YCaH$_{18}$ were calculated to be at or within 12~meV/atom of the convex hull when $S_{\text{vib.}}$ was considered between 0-500~K (Figures~\ref{fig:YCaH_convex_zoom_vib_100GPa}(a-d) and Table~\ref{tab:YCaH_ConvexHullDistances_100GPa}).  Since we identified three nearly isoenthalpic $Cmcm$ based YCaH$_8$ phases at this pressure, the effect of $S_{\text{config.}}$ was considered, and including it placed this stoichiometry within $\sim$3~meV/atom of the convex hull at 3500~K (Figure~\ref{fig:YCaH_convex_zoom_S-select_100GPa}(j)).

\item At 150 and 200~GPa, $Fd\bar{3}m$ YCaH$_{12}$ lays on the convex hull at and below 1500~K when $S_{\text{vib.}}$ and the ZPE were included (Figures~\ref{fig:YCaH_convex_zoom_vib_150GPa}(a-f) and \ref{fig:YCaH_convex_zoom_vib_200GPa}(a-f) and Tables~\ref{tab:YCaH_ConvexHullDistances_150GPa} and \ref{tab:YCaH_ConvexHullDistances_200GPa}). $S_{\text{config.}}$ brought the YCaH$_{12}$ stoichiometry to the hull at all of the temperatures considered at both pressures (Figures~\ref{fig:YCaH_convex_zoom_S-select_150GPa}(a-j) and \ref{fig:YCaH_convex_zoom_S-select_200GPa}(a-j)), while at 200~GPa, the YCaH$_8$ composition also graced the convex hull at and above 1000~K (Figures~\ref{fig:YCaH_convex_zoom_S-select_200GPa}(e-j)).

\item At 250~GPa, $Fd\bar{3}m$ YCaH$_{12}$ was thermodynamically stable to 2000~K considering $S_{\text{vib.}}$ (Figures~\ref{fig:YCaH_convex_zoom_vib_250GPa}(a-g) and Table~\ref{tab:YCaH_ConvexHullDistances_250GPa}), but between 2500-3500~K $Cmmm$ YCaH$_8$ lays on the convex hull (Figures~\ref{fig:YCaH_convex_zoom_vib_250GPa}(h-j)). $S_{\text{config.}}$ stabilized YCaH$_{12}$ at all temperatures (Figures~\ref{fig:YCaH_convex_zoom_S-select_250GPa}(a-j)), while YCaH$_8$ comprised the convex hull at and above 1000~K (Figures~\ref{fig:YCaH_convex_zoom_S-select_250GPa}(e-j)).

\item $Fd\bar{3}m$ YCaH$_{12}$ was on the 300~GPa convex hull at all considered temperatures (Figures~\ref{fig:YCaH_convex_zoom_vib_300GPa}(a-j) and Table~\ref{tab:YCaH_ConvexHullDistances_300GPa}), and it was joined by $Cmmm$ YCaH$_8$ at 3000~K (Figure~\ref{fig:YCaH_convex_zoom_vib_300GPa}(i)). Inclusion of $S_{\text{config.}}$ brought the YCaH$_8$ composition to the convex hull alongside YCaH$_{12}$ at and above 1000~K (Figures~\ref{fig:YCaH_convex_zoom_S-select_300GPa}(e-j)).
\end{itemize}

\subsection{Electronic Structure and Superconductivity}
\begin{figure}[!htb]
    \centering
    \includegraphics[width=16.5 cm]{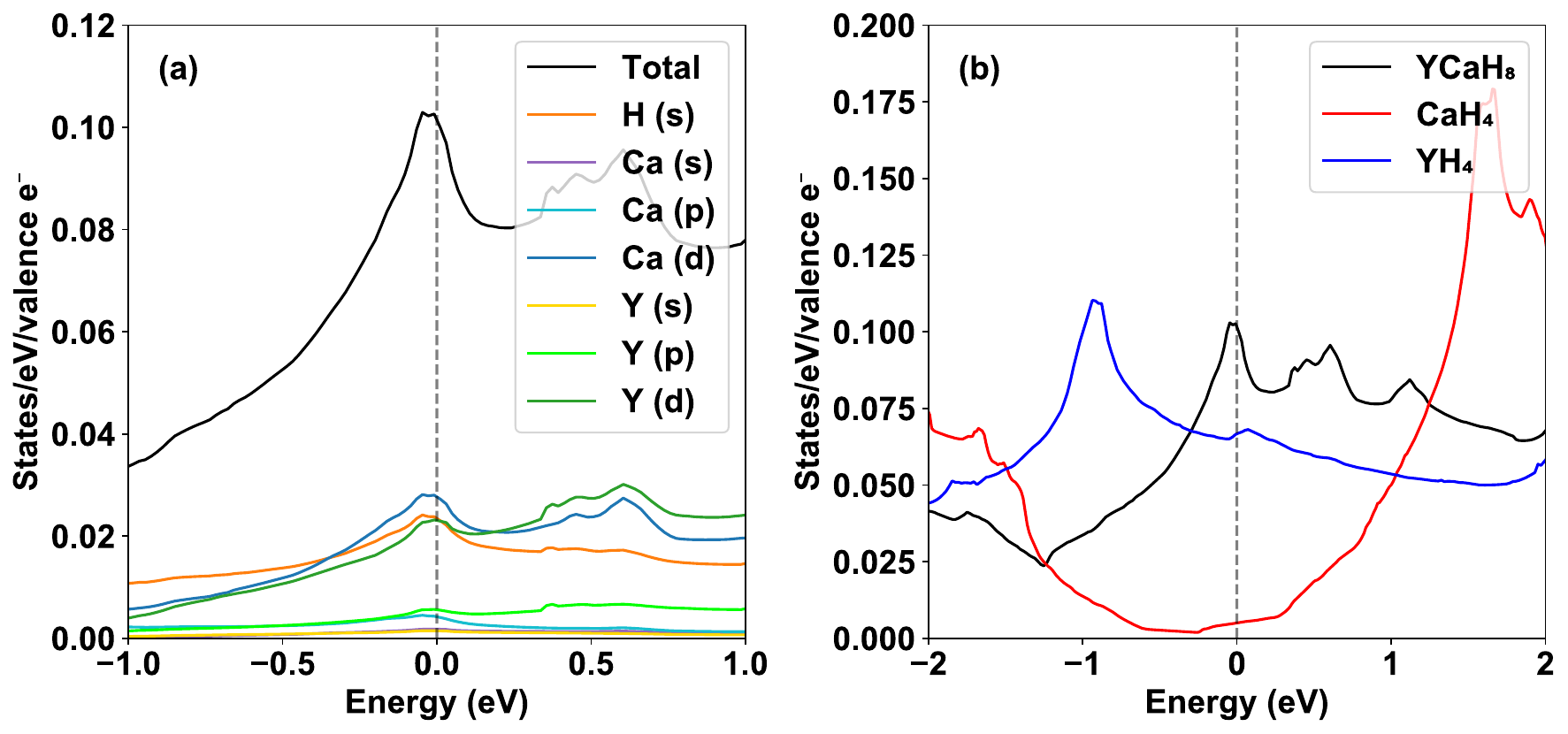}
    \caption{(a) The total and orbital-projected density of states (DOS) of $P4/mmm$ YCaH$_{8}$ at 150~GPa. (b) The total DOS of $P4/mmm$ YCaH$_{8}$ (black), $I4/mmm$ CaH$_{4}$ (red), and $I4/mmm$ YH$_{4}$ (blue) at 150~GPa. The Fermi energy is set to 0~eV. The total and projected DOS are normalized by the total number of valence electrons, taken as Ca (2~$e$), Y (3~$e$), and H (1~$e$).}\label{fig:YCaH8_P4.mmm_dos}
\end{figure}

The PBE electronic band structures and DOS plots (Figures~\ref{fig:YCaH8_Cmmm_BandDOS}-\ref{fig:YCaH20_P2_BandDOS}) show that most of the dynamically stable YCaH$_{n}$ phases, and in particular all of the ordered arrangements of the YCaH$_8$ and YCaH$_{12}$ superhydrides considered herein, are metallic. Though charge is donated from the electropositive metal atoms to the hydrogenic lattice, whose 1s-states are abundant near $E_\text{F}$, substantial metal character is evident as well, in-line with the Bader charges of +0.88 to +1.14 for Ca, and +1.28 to +1.60 for Y, suggesting incomplete charge transfer.
Because Ca undergoes an electronic 4s$\rightarrow$3d transition under pressure~\cite{Maksimov:2005a}, substantial Ca 3d character is observed around $E_\text{F}$, mirroring previous computational results obtained for CaH$_4$~\cite{Mishra:2018a}. Another key contribution to the metallicity can be attributed to the Y 4d states. These features are illustrated for  $P4/mmm$ YCaH$_8$ (Figure~\ref{fig:YCaH8_P4.mmm_dos}(a)); however they are universal for all of the metallic ternary hydride phases.

Assuming full electron transfer from the electropositive element to the hydrogenic lattice, the formula for $I4/mmm$ CaH$_4$ can be written as [Ca$^{2+}$][2H$^-$]$\cdot$H$_2$; this phase is therefore not a metal at ambient pressure. Within PBE, the band gap closes by 60~GPa due to pressure induced band broadening of the valence and conduction bands, rendering CaH$_4$ weakly metallic~\cite{Mishra:2018a}. Yttrium possesses one more valence electron than calcium, so assuming full oxidation, the formula of its tetrahydride can be written as [Y$^{3+}$][2H$^-$]$\cdot$H$_2^{-}$. The half filling of the H$_2$ $\sigma^*$-band stretches the H-H bond, and induces metallicity~\cite{Bi:2021a}. At 150~GPa a peak is evident in the DOS above (below) $E_\text{F}$ in  $I4/mmm$ CaH$_4$ (YH$_4$), and adding (subtracting) 0.5$e$ per formula unit places $E_\text{F}$ on top of this peak in $P4/mmm$ YCaH$_8$ at 150~GPa  (Figure~\ref{fig:YCaH8_P4.mmm_dos}(b)), and also at higher pressures  (Figure~\ref{fig:YCaH8_P4.mmm_BandDOS}). These results suggest that $P4/mmm$ YCaH$_8$ might have a higher \tc\ than the binary hydrides from which it is derived. Such peaks in the DOS often result from the presence of van Hove singularities at or near $E_\text{F}$, and have been noted in other high-pressure hydrides~\cite{quan2016van,Zurek:2020g}. However, $E_\text{F}$ did not lie on a peak in the DOS in all of the $I4/mmm$-based YCaH$_8$ phases (Figures~\ref{fig:YCaH8_Cmmm_BandDOS}, \ref{fig:YCaH8_Imma_BandDOS}, and \ref{fig:YCaH8_C2.m-alpha_BandDOS}), suggesting the metal arrangement may also impact \tc . The structural renormalization that can occur upon inclusion of quantum nuclear affects~\cite{Zurek:2025c} can in some cases have a substantial influence on the computed \tc s; and it will be the focus of our future studies.

Due to the computational expense associated with performing EPC calculations, the {\tc}s of select YCaH$_{n}$ phases were estimated using TcESTIME~\cite{Novoa:2025a} with the RS5~\cite{Belli:2025a} fit, which improves upon the original ELF-based fit~\cite{Belli:2021a} by incorporating the molecularity index~\cite{diMauro:2024a}, including ternary hydrides in the dataset, and achieving a root mean squared error of $\sim$40~K. The {\tc}s of the ordered YCaH$_{8}$ phases (Table~\ref{tab:YCaH8_tcestime}), specifically those derived from $I4/mmm$ MH$_{4}$, were predicted to be within a few Kelvin of one another at 300~GPa, but the \tc\ of the $P4/mmm$ phase was predicted to be significantly lower than those of the other colorings below 270~GPa. Meanwhile, the $Cmcm$ MH$_{4}$-derived $Amm2$ YCaH$_{8}$ phase was predicted to have a lower {\tc} at 300~GPa than the $I4/mmm$ derived phases. Similarly, this fit suggested that the ordered YCaH$_{12}$ phases possess similar {\tc}s from 200-300~GPa, with the exception of $Amm2$ at 200~GPa, primarily due to its lower networking value (Table~\ref{tab:YCaH12_tcestime}). Additionally (Tables~\ref{tab:YCaH8_tcestime}-\ref{tab:YCaH20_tcestime}), many stable and metastable metallic YCaH$_{n}$ phases are predicted to be high-{\tc} superconductors with this ELF-based fit.

\begin{table}[!h]
    \centering

    \caption{Superconducting properties of select YCaH$_{n}$ phases. The {\tc}s were calculated via the Allen-Dynes modified McMillan formula (ADM) with strong-coupling and shape corrections and $\mu^{*}$ = 0.10, as well as the numerical solutions of the Eliashberg equations (Eliash.)\ with $\mu^{*}$ = 0.10 and 0.13, and with the RS5 fit.}

    \begin{tabular}{M{0.45in}M{0.7in}M{0.7in}M{0.5in}M{0.75in}M{0.75in}M{0.8in}M{0.50in}}
    \toprule\toprule
    \textbf{System} & \textbf{Space Group} & \textbf{Pressure (GPa)} & \textbf{$\pmb{\lambda}$} & \textbf{$\pmb{\omega_{\text{ln}}}$ (K)} & \textbf{$\pmb{T_{c}}$ (K) (ADM)} & \textbf{$\pmb{T_{c}}$ (K) (Eliash.)} & \textbf{$\pmb{T_{c}}$ (K) (RS5)} \\
    \midrule

    \multirow{4}{*}{YCaH$_{8}$}  & \multirow{2}{*}{$P4/mmm$} & 150 & 1.49 & 925.9  & 120.2 & 139-149 & 36 \\
                                 &                           & 180 & 1.38 & 1024.4 & 120.9 & 139-149 & 30 \\\cline{2-8}
                                 & \multirow{2}{*}{$Cmmm$}   & 180 & 1.64 & 859.6  & 124.8 & 161-170 & 115 \\
                                 &                           & 200 & 1.48 & 1061.5 & 136.7 & 160-170 & 111 \\
    \hline
    \multirow{4}{*}{YCaH$_{12}$} & $Fd\bar{3}m$              & 200 & 1.98 & 1297.8 & 220.4$^{a}$ & 241-253$^{b}$ & 169 \\
                                 & $Pm\bar{3}m$              & 200 & 1.50 & 1273.3 & 166.2$^{c,d}$ & 181-192$^{e}$ & 169 \\
                                 & $P4/nmm$                  & 200 & 1.67 & 1269.7 & 189.2 & 199-210 & 168 \\
                                 & $P3m1$                    & 200 & 0.96 & 1313.6 &  91.3 & 96-105  & 169 \\
    \hline
    YCaH$_{18}$                  & $Pmmn$-$\alpha$           & 200 & 1.73 & 1047.6 & 159.6 & 182-193 & 64 \\

    \bottomrule
    \end{tabular}

    \raggedright
    $^{a}$ 226~K [Ref.~\cite{Liang:2019a}], 220~K [Ref.~\cite{Zhao:2025a}] \\
    $^{b}$ 243-258~K [Ref.~\cite{Liang:2019a}] \\
    $^{c}$ 215~K [Ref.~\cite{Liang:2019a}] \\
    $^{d}$ 222~K [Ref.~\cite{Xie:2019a}] \\
    $^{e}$ 233-248~K [Ref.~\cite{Liang:2019a}] \\

\label{tab:epc}
\end{table}

\begin{figure}[!htb]
    \centering
    \includegraphics[width=16.5 cm]{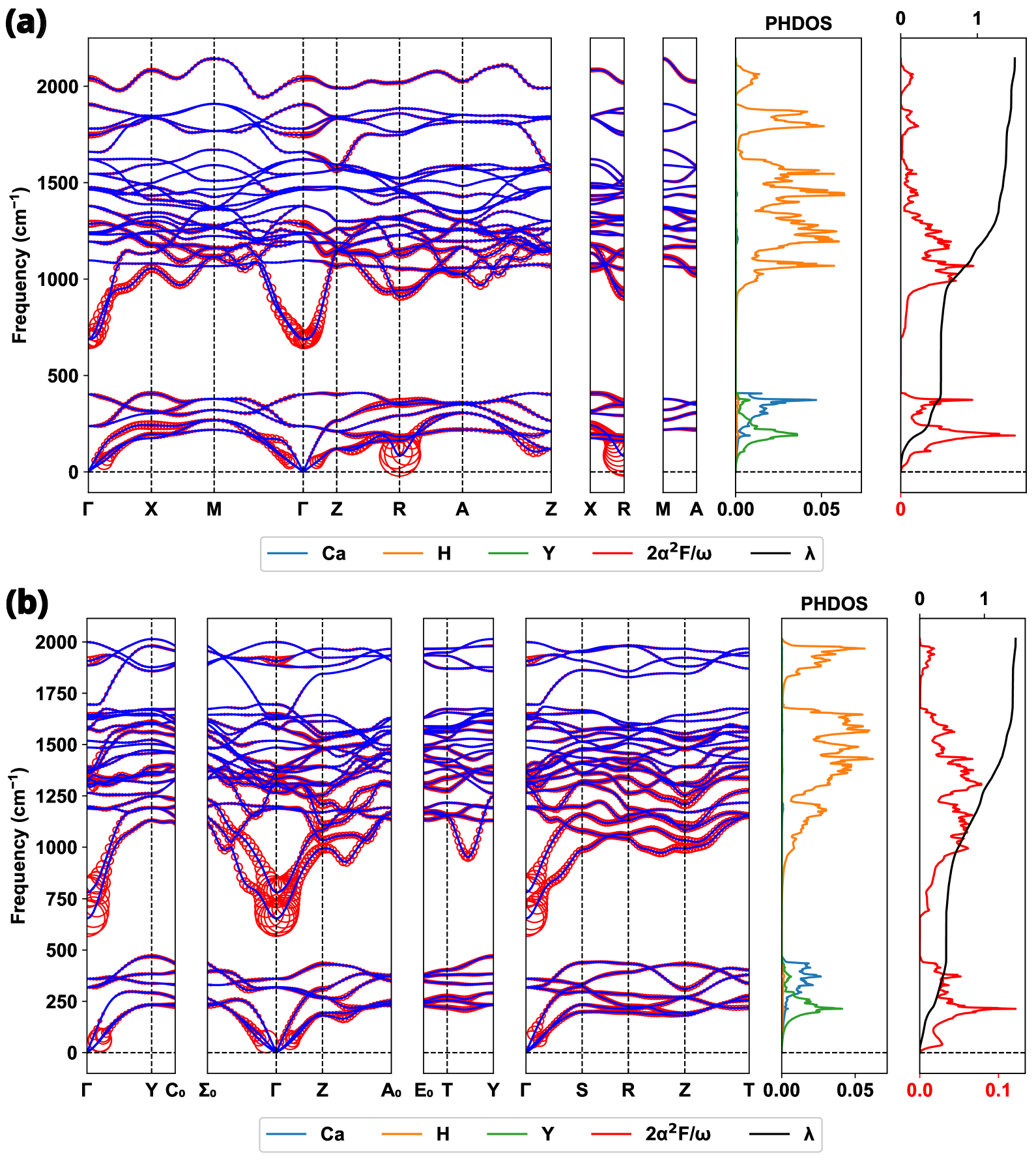}
    \caption{Phonon dispersion curves, projected phonon DOS (PHDOS), the Eliashberg spectral function scaled by frequency ($2\alpha^{2}F/\omega$), and the integrated EPC constant ($\lambda$) of (a) $P4/mmm$ and (b) $Cmmm$ YCaH$_{8}$ at 150 and 200~GPa, respectively. The radii of the red bubbles are proportional to the strength of the electron-phonon coupling constant ($\lambda_{\nu \textbf{q}}$) for the vibrational mode at $\nu$ and wavevector $\textbf{q}$.}\label{fig:YCaH8_epc}
\end{figure}

To confirm the superconducting properties of select stable and metastable YCaH$_{n}$ structures that have the potential of hosting high-{\tc} superconductivity, DFPT calculations were performed.  The resulting EPC parameter, $\lambda$, the logarithmic average of the phonon frequencies, $\omega_\text{ln}$, and the Allen-Dynes modified McMillan (ADM) {\tc}s (including the strong-coupling and shape corrections) are tabulated in Tables~\ref{tab:epc} and \ref{tab:epc_data}, as are results obtained by numerically solving the Eliashberg equations. 
At 150~GPa, the Eliashberg {\tc} of $P4/mmm$ YCaH$_{8}$ was estimated to be 139-149~K, which is $\sim$1.6 times the value computed for $I4/mmm$ YH$_{4}$ at these same conditions (Table~\ref{tab:epc_binary_results}). The {\tc} for $I4/mmm$ CaH$_{4}$ was not calculated, as it is either weakly metallic or semiconducting at high pressures.\cite{Mishra:2018a}. Though the substitution of half of the Y atoms by Ca decreased $\omega_\text{ln}$ somewhat, it increased $\lambda$ substantially from 0.95 to 1.49, resulting in the increase of \tc . A doubly degenerate mode at $\sim$690~cm$^{-1}$ at $\Gamma$ (Figure~\ref{fig:YCaH8_epc}a) was found to possess a particularly strong mode-resolved EPC constant ($\lambda_{\nu \textbf{q}}$). Visualization of the vibration associated with this soft mode showed that it corresponded to a coupled libration-stretching mode of the H$_2$ molecules, resulting in the formation of quadrilaterals involving the H$_{a}$ and H$_{b}$ atoms, similar to motions with large EPC described previously for other MH$_{4}$ structures~\cite{Bi:2021a}.  Another soft phonon mode with a considerable $\lambda_{\nu \textbf{q}}$ is present at the $R$ high symmetry point at $\sim$81~cm$^{-1}$; visualization revealed it to be a breathing mode of the hydrogenic dodecahedral cage surrounding the Ca atoms. The relatively flat phonon band at $\sim$1100~cm$^{-1}$, which leads to a high phonon density of states (PHDOS) and a peak in the Eliashberg spectral function in this region, consists of translations of the H$_{2}$s, specifically those side-on to the Ca and Y atoms, along the $c$ lattice vector.  At 180~GPa, a decrease in $\lambda$ in $P4/mmm$ YCaH$_8$ is offset by an increase in $\omega_\text{ln}$ resulting in a nearly equivalent \tc\ (Table \ref{tab:epc}), and the soft mode at $R$ is no longer present.  At higher pressures, the soft phonon mode at $\Gamma$ hardens (Figures~\ref{fig:YCaH8_P4.mmm_EPCPlot_200GPa}-\ref{fig:YCaH8_P4.mmm_EPCPlot_300GPa}) and the electronic DOS at $E_\text{F}$ decreases (Table~\ref{tab:epc_data}), leading to lower EPCs and Eliashberg {\tc}s. Between 150-300~GPa, no vibrational modes exceed 2500~cm$^{-1}$, in-line with the H-H distances that measure at least 0.90~\AA. At 140~GPa, $P4/mmm$ YCaH$_8$ becomes dynamically unstable with an imaginary mode at the $R$ point (Figure~\ref{fig:YCaH8_P4.mmm_EPCPlot_140GPa}); however, this same structure has been reported to be dynamically stable at 100~GPa with an ADM {\tc} of 90~K~\cite{Zhao:2025a}.

At 200~GPa $Cmmm$ YCaH$_{8}$ possesses two soft phonon modes at $\Gamma$, one at $\sim$690~cm$^{-1}$ and the other at $\sim$780~cm$^{-1}$, with large $\lambda_{\nu \textbf{q}}$s (Figure~\ref{fig:YCaH8_epc}(b)). Visualization shows that both of these are coupled libration-stretches of the H$_2$ molecules, but the first involves H$_2$s oriented side-on to the Y atom, while the other mode involves H$_2$s that interact with Ca. At 180~GPa the \tc\ of $Cmmm$ YCaH$_8$ was predicted to be $\sim$20~K higher than that of $P4/mmm$ YCaH$_8$ within the harmonic approximation (Table~\ref{tab:epc}), further highlighting the potential of metal atom substitution in enhancing the superconducting properties of multi-element hydrides. At 200~GPa, the Eliashberg {\tc} of the $Cmmm$ symmetry phase is approximately the same as the one computed at 180~GPa. Further mirroring the results obtained for $P4/mmm$ YCaH$_8$, the {\tc} of the $Cmmm$ phase decreases markedly at higher pressures (Table~\ref{tab:epc_data}).  Additionally, we computed the {\tc} of the $Cmcm$ MH$_{4}$-based $Amm2$ YCaH$_{8}$ structure (Table~\ref{tab:epc_data}) to compare with those derived from $I4/mmm$ MH$_{4}$. The ADM {\tc} of $Amm2$ (59~K) at 300~GPa was found to not only be lower than the one calculated for $P4/mmm$ (83~K) and $Cmmm$ (70~K) YCaH$_{8}$, but also lower than the ADM \tc\ of $I4/mmm$ YH$_{4}$ (65~K); however, it has a higher {\tc} than $Cmcm$ CaH$_{4}$ (40~K)~\cite{Shao:2019a} at the same pressure.

Here, we note some discrepancies between the {\tc}s predicted by the RS5 fit (Tables~\ref{tab:YCaH8_tcestime}-\ref{tab:YCaH20_tcestime}, and Table~\ref{tab:epc}) and those computed from first principles (Table~\ref{tab:epc_data} and Table~\ref{tab:epc}) for the $P4/mmm$ and  $Cmmm$ YCaH$_8$ phases. With the exception of a few pressures where the \tc\ predicted by the RS5 fit either overestimated the DFT value or fell within the error expected for the fit (e.g.\ at 300~GPa for $P4/mmm$, or 250 and 300~GPa for $Cmmm$), the first-principles \tc\ is  significantly underestimated. The error is particularly large for the $P4/mmm$ phase at 150~GPa with the difference being as high as $\sim$109-119~K, almost triple the mean-squared error of the RS5 fit, when compared against the Eliashberg values. 
This discrepancy can be attributed to the high molecularity index computed for $P4/mmm$ at and below 260~GPa resulting from the short H-H distances ($\le$1.02~\AA{}) for the H$_2$s that surround Ca side-on. For binary hydrides the presence of such motifs quenches superconductivity, but in the ternaries studied here the H$_2$ molecules surrounding Y have longer bonds due to donation from the metal atom to the H$_2$ $\sigma^*$-bands. We speculate that including compounds possessing both short and long H-H distances (such as $P4/mmm$ YCaH$_8$ at lower pressures) into the ELF-based fit might improve \tc\ predictions for multi-component MH$_4$ systems.

Let us now turn to the superconducting properties of the YCaH$_{12}$ phases. At 200~GPa, both CaH$_{6}$~\cite{Wang:2012a,Ma:2022b,Li:CaH6} and YH$_{6}$~\cite{Wang:2022a,Troyan:2021a,Kong:2021a} are predicted (and measured) to have high {\tc}s of 223-233~K and 246-258~K, respectively (Table~\ref{tab:epc_data_binary_hydrides}). 
The DOS at $E_\text{F}$ for all of the YCaH$_{12}$ phases were calculated to be larger than in CaH$_{6}$, but smaller than in YH$_{6}$ at this pressure. We therefore wondered if the YCaH$_{12}$ phases could have {\tc}s that are higher than that of CaH$_{6}$ but lower than for YH$_{6}$? To answer this question, we performed EPC calculations on select YCaH$_{12}$ phases.
Although the superconducting properties of $Fd\bar{3}m$\cite{Liang:2019a,Zhao:2025a} and $Pm\bar{3}m$\cite{Liang:2019a,Xie:2019a} YCaH$_{12}$ have been studied before, we chose to perform EPC calculations to verify the consistency of our results. At 200~GPa, the ADM and Eliashberg {\tc}s of $Fd\bar{3}m$ are estimated to be 220~K and 241-253~K, respectively, and as can be seen in Tables~\ref{tab:epc} and ~\ref{tab:epc_data}, they are consistent with some previously reported {\tc}s~\cite{Liang:2019a,Zhao:2025a}, albeit with higher $\lambda$s and lower $\omega_{\text{ln}}$s. $Fd\bar{3}m$ YCaH$_{12}$ was estimated to have a slightly lower Eliashberg {\tc} than YH$_6$ by $\sim$5~K, while having a higher {\tc} than CaH$_6$ by $\sim$18~K. The {\tc} of $Fd\bar{3}m$ YCaH$_{12}$ increased at lower pressures, reaching a maximum Eliashberg value of 254-265~K at 160~GPa before becoming dynamically unstable at 150~GPa (Figure~\ref{fig:YCaH12_Fd-3m_EPCPlot_150GPa}). Meanwhile, both $\lambda$ and the Eliashberg \tc\ of $Pm\bar{3}m$ YCaH$_{12}$ were found to be significantly lower than previously reported results~\cite{Liang:2019a,Xie:2019a}. We must also note that we used the same $\pmb{k}$- and $\pmb{q}$-point meshes as Liang \emph{et al.}~\cite{Liang:2019a} for computing the EPC matrix elements. Interestingly, by using courser $\pmb{k}$- and $\pmb{q}$-point meshes, we obtained similar $\lambda$s and {\tc}s to Liang \emph{et al.}, 
indicating that $Pm\bar{3}m$ is sensitive to the choice of computational parameters.

Two other YCaH$_{12}$ phases we opted to perform EPC calculations on are $P4/nmm$ and $P3m1$. As tabulated in Table~\ref{tab:epc}, the {\tc}s of $P4/nmm$ were computed to be lower than that of $Fd\bar{3}m$ YCaH$_{12}$ and higher than that of $Pm\bar{3}m$, while $P3m1$ was estimated to have a significantly smaller {\tc} than all three. This drastically contrasts the {\tc}s predicted by the RS5 fit as the Eliashberg {\tc}s 
are not similar (Table~\ref{tab:epc}). Furthermore, their molecularity indices, networking values, and electronic DOS at $E_{F}$, specifically the percentage projected onto the hydrogen atoms, were also found to be similar 
(Table~\ref{tab:YCaH12_tcestime}). Moreover, and as seen in Table~\ref{tab:epc_data}, the electronic DOS at $E_{F}$ of these four phases are within 0.02~states/eV/FU of each other at 200~GPa, while the largest difference in $\omega_{\text{ln}}$ is $\sim$44~K between $P3m1$ and $P4/nmm$. Despite these similarities, the primary difference between these phases is their symmetries, where a lower symmetry ($P3m1$) will have less degeneracies compared to a higher symmetry one ($Fd\bar{3}m$). 

Lastly, we performed EPC calculations on $Pmmn$-$\alpha$ YCaH$_{18}$ as it lays within 10~meV/atom of the cold convex hull at 200-300~GPa. The Eliashberg {\tc} of this phase was calculated to lie between 182-193~K at 200~GPa, which is lower than the value computed for $Pnma$ YH$_{9}$ at this pressure (227-243~K)~\cite{Du:2023a}. 
This is another system where the RS5 fit significantly underestimated the \tc\ at 200~GPa (Table \ref{tab:epc}). We hypothesize the discrepancy to be a result of the large molecularity index that results from a short H-H distance of 0.87~{\AA} at this pressure.

\section{Conclusions}
The atomic radius of Ca is somewhat smaller than that of Y, but the opposite is true for the most common oxidation states of these elements, Ca$^{2+}$ and Y$^{3+}$. In the binary high pressure hydrides, these differences manifest in the compositions of the most stable superhydrides and their thermodynamic stabilities. While both $I4/mmm$ and $Cmcm$ CaH$_4$ and YH$_4$, and $Im\bar{3}m$ CaH$_6$ and YH$_6$ lie on the convex hull in their respective pressure ranges, thermodynamically stable $P6_{3}/mmc$ YH$_9$ and $Fm\bar{3}m$ YH$_{10}$ have been predicted to possess clathrate-like hydrogenic lattices, while the preferred crystal lattices of metastable CaH$_9$ and CaH$_{10}$ have reduced dimensionalities. 
We therefore wondered how these chemistries might impact the properties of high-pressure superhydrides that contain both metal elements?

To determine the structures and properties of the phases that may form if a Ca-Y alloy is reacted with hydrogen under pressure, density functional theory (DFT) calculations were performed on YCaH$_n$ ($n = 8-20$) phases between 100-300~GPa. Notable phases that emerged in our DFT studies possessed  YCaH$_8$, YCaH$_{12}$, YCaH$_{18}$, and YCaH$_{20}$ stoichiometries, and they could be derived from the structures of the previously reported  $I4/mmm$ and $Cmcm$ MH$_{4}$, $Im\bar{3}m$ MH$_{6}$, $P6_{3}/mmc$ MH$_{9}$, and $Fm\bar{3}m$ MH$_{10}$ superhydrides, respectively. For the YCaH$_8$ and YCaH$_{12}$ stoichiometries, numerous phases that differed in the ordering of the metal atoms were found to be enthalpically competitive, hinting at potential metal disorder and stabilization via configurational entropy. Though various YCaH$_{18}$ and YCaH$_{20}$ configurations possessing different metal decorations were nearly isoenthalpic, only a single phase of each stoichiometry was found to be dynamically stable within the harmonic approximation, in-line with the differences in the stabilities and structures previously noted for their respective binary hydrides.

The thermodynamic stabilities of YCaH$_n$ phases were further assessed by including the zero-point energy (ZPE), vibrational entropy, and the configurational entropy (only for $n=8, 12$). Between 200-300~GPa, YCaH$_{8}$ lies on the convex hull at and above 1000~K with ZPE, vibrational entropy, and configurational entropy. YCaH$_{12}$ was found to lie on the 0~K convex hulls at 150-300~GPa when considering only the enthalpy and ZPE, but it became thermodynamically unstable above 1500~K (at 150 and 200~GPa) and 2000~K (at 250~GPa) when vibrational entropy was included. With configurational entropy, however, YCaH$_{12}$ lies on the convex hull between 150-300~GPa at all considered temperatures, highlighting the importance of disorder in determining phase stability.

The considered 
YCaH$_{8}$ phases, specifically those that were selected for electron-phonon coupling calculations, were found to possess a higher density of states (DOS) at the Fermi level ($E_\text{F}$) than their parent binary hydrides, resulting in predicted enhanced superconducting temperatures (Eliashberg \tc s) of 139-149~K for the $P4/mmm$ and 161-170~K for the $Cmmm$ symmetry phases at 180~GPa. Meanwhile, the predicted \tc s of the studied YCaH$_{12}$ phases spanned a broad range, 96-253~K at 200~GPa, with higher symmetry structures generally possessing increased \tc . Thus, while tuning the electronic configuration by alloying Y and Ca to reach a peak in the DOS can increase \tc\ in the MH$_4$-based hydrides, within those based on MH$_6$ the effect is not pronounced as the 
$E_\text{F}$ will be shifted to a region with lower DOS in YH$_{6}$ upon substituting Y with an equimolar number of Ca atoms, and although the opposite occurs when substituting Ca with an equimolar number of Y atoms in CaH$_{6}$, the DOS at $E_\text{F}$ in the resulting YCaH$_{12}$ phases will be lower than YH$_{6}$.

Further work is suggested to study the effects of anharmonicity on the structural characteristics and dynamic stability of the various ordered phases considered herein, using descriptors~\cite{Zurek:2025c} and machine-learning accelerated workflows~\cite{Zurek:2024g}. Moreover, our studies highlight that the presence of inequivalent H$_2$ units with differing bond lengths can result in large deviations in \tc s predicted by models based upon the networking properties of the Electron Localization Function, which could be re-fitted based upon the distance of the longest H-H bond for the molecularity index calculations~\cite{Belli:2025a}.

\section{Acknowledgments}
Funding for this research was provided by the U.S. Department of Energy, National Nuclear Security Administration, through the Chicago-DOE Alliance Center (CDAC) under Cooperative Agreement DE-NA0004153, the National Science Foundation under award DMR-2136038, and the U.S. Department of Energy, Office of Science, Fusion Energy Sciences
funding the award entitled ``High Energy Density Quantum Matter,'' Award No.\ DE-SC0020340. Calculations were performed at the Center for Computational Research at SUNY Buffalo (\url{https://hdl.handle.net/10477/79221}).

\section{Supporting Information Available}
The Supporting Information is available free of charge on the publication website, \url{https://pubs.acs.org/}. It includes more informative computational details, ternary convex hulls, relative enthalpies, equation of states, convex hull distances, formation enthalpies and Gibbs Free energies, phonon band structures and density of states, images of select phases, the electronic band structures and density of states, {\tc} prediction results using the RS5 fit, computational parameters used in the EPC calculations, superconducting properties of select phases, and structural parameters. The optimized structures are available on the NOMAD~\cite{Scheidgen:2023a} repository (DOI: \href{https://doi.org/10.17172/nomad.y5wv-afh8}{10.17172/nomad.y5wv-afh8}).

\section{Notes}
The authors declare no competing financial interest.

\bibliography{YCaH,NSF-clean}

\clearpage
\newpage

\section{For Table of Contents Only}
\begin{figure}[!ht]
    \centering
    \includegraphics[width=3.25 in]{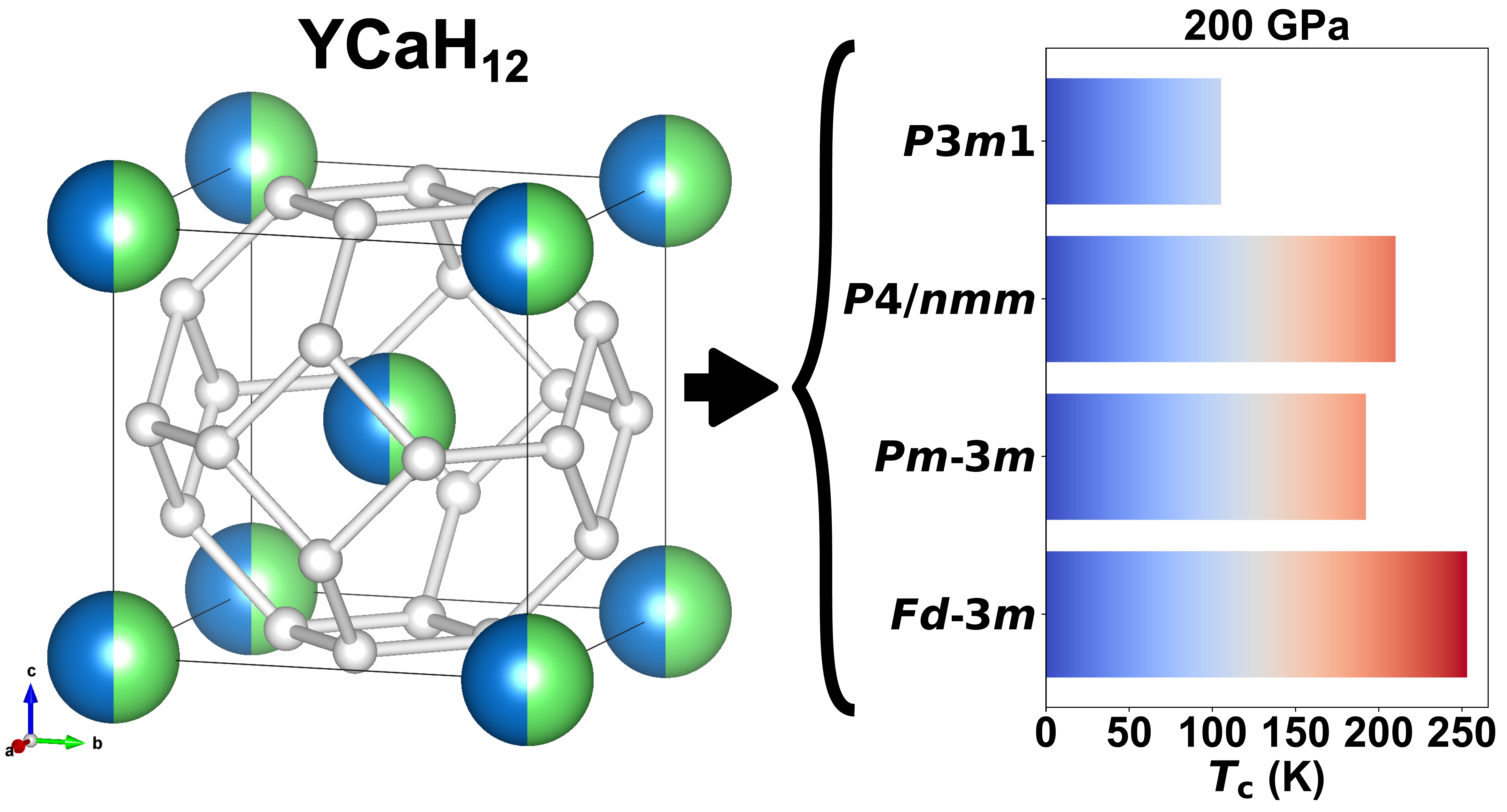}
\end{figure}

\end{document}